\documentclass{aastex63}

\usepackage[T1]{fontenc}  
\usepackage{amsmath}      
\usepackage{mathtools}    
\usepackage{tensor}       
\hypersetup{colorlinks=true,linkcolor=black,citecolor=black,filecolor=black,urlcolor=black}

\makeatletter
\newcommand{\extref}[1]{\textup{\tagform@{#1}}}
\makeatother

\newcommand{\ee}{\mathrm{e}}
\newcommand{\dd}{\mathrm{d}}
\newcommand{\dr}{\dd r}
\newcommand{\dth}{\dd\theta}
\newcommand{\dph}{\dd\phi}
\newcommand{\dtau}{\dd\tau}
\newcommand{\pp}{\partial}
\newcommand{\pz}{\pp z}
\DeclarePairedDelimiter{\paren}{\lparen}{\rparen}
\DeclarePairedDelimiter{\abs}{\lvert}{\rvert}
\DeclarePairedDelimiter{\ave}{\langle}{\rangle}

\newcommand{\xmin}{x_\mathrm{min}}
\newcommand{\xmax}{x_\mathrm{max}}
\newcommand{\rmin}{r_\mathrm{min}}
\newcommand{\rmax}{r_\mathrm{max}}
\newcommand{\thetamin}{\theta_\mathrm{min}}
\newcommand{\thetamax}{\theta_\mathrm{max}}
\newcommand{\thetac}{\theta_\mathrm{c}}
\newcommand{\rhor}{r_\mathrm{hor}}
\newcommand{\pgas}{p_\mathrm{gas}}
\newcommand{\pgasmin}{p_\mathrm{gas,min}}
\newcommand{\pmag}{p_\mathrm{mag}}
\newcommand{\mdot}{\dot{M}}
\newcommand{\mdotin}{\dot{M}_\mathrm{in}}
\newcommand{\mdotout}{\dot{M}_\mathrm{out}}
\newcommand{\edot}{\dot{E}}
\newcommand{\edotmag}{\dot{E}_\mathrm{mag}}
\newcommand{\jdot}{\dot{J}}
\newcommand{\tvisc}{t_\mathrm{visc}}
\newcommand{\rvisc}{r_\mathrm{visc}}
\newcommand{\qconv}{Q_\mathrm{conv}}
\newcommand{\qcirc}{Q_\mathrm{circ}}
\newcommand{\be}{\mathrm{Be}}
\newcommand{\msun}{M_\odot}
\newcommand{\second}{\mathrm{s}}
\newcommand{\hour}{\mathrm{h}}
\newcommand{\ghz}{\mathrm{GHz}}
\newcommand{\jy}{\mathrm{Jy}}

\newcommand{\athena}{\texttt{Athena++}}
\newcommand{\ibothros}{\texttt{ibothros}}

\shorttitle{Radiatively Inefficient Accretion}
\shortauthors{C.~J.~White, E.~Quataert, C.~F.~Gammie}

\begin{document}

\title{The Structure of Radiatively Inefficient Black Hole Accretion Flows}
\author{Christopher~J.~White}
\affiliation{Kavli Institute for Theoretical Physics, University of California Santa Barbara, Kohn Hall, Santa Barbara, CA 93107, USA}
\author[0000-0001-9185-5044]{Eliot~Quataert}
\affiliation{Department of Astronomy, University of California Berkeley, 501 Campbell Hall, Berkeley, CA 94720, USA}
\author[0000-0001-7451-8935]{Charles~F.~Gammie}
\affiliation{Department of Astronomy and Department of Physics, University of Illinois, Urbana, IL 61801, USA}

\begin{abstract}
  We run three long-timescale general-relativistic magnetohydrodynamic simulations of radiatively inefficient accretion flows onto non-rotating black holes. Our aim is to achieve steady-state behavior out to large radii and understand the resulting flow structure. A simulation with adiabatic index $\Gamma = 4/3$ and small initial alternating poloidal magnetic field loops is run to a time of $440{,}000\ GM/c^3$, reaching inflow equilibrium inside a radius of $370\ GM/c^2$. Variations with larger alternating field loops and with $\Gamma = 5/3$ are run to $220{,}000\ GM/c^3$, attaining equilibrium out to $170\ GM/c^2$ and $440\ GM/c^2$. There is no universal self-similar behavior obtained at radii in inflow equilibrium:\ the $\Gamma = 5/3$ simulation shows a radial density profile with power law index ranging from $-1$ in the inner regions to $-1/2$ in the outer regions, while the others have a power-law slope ranging from $-1/2$ to close to $-2$. Both simulations with small field loops reach a state with polar inflow of matter, while the more ordered initial field has polar outflows. However, unbound outflows remove only a factor of order unity of the inflowing material over a factor of $\sim 300$ in radius. Our results suggest that the dynamics of radiatively inefficient accretion flows are sensitive to how the flow is fed from larger radii, and may differ appreciably in different astrophysical systems. Millimeter images appropriate for Sgr~A* are qualitatively (but not quantitatively) similar in all simulations, with a prominent asymmetric image due to Doppler boosting.
\end{abstract}

\section{Introduction}
\label{sec:introduction}

A common mode of accretion onto black holes is that of a radiatively inefficient accretion flow (RIAF). Such systems have nonzero net angular momentum, as opposed to classical Bondi accretion \citep{Bondi1952}, but because of inefficient cooling they are dynamically hot, thick disks \citep{Ichimaru1977,Rees1982,Narayan1995}, unlike standard thin disk models \citep{Shakura1973}. RIAFs are consistent with a number of low-luminosity active galactic nuclei, including Sgr~A* \citep{Narayan1998}.

Simulations of black hole accretion on horizon scales, where general relativity (GR) cannot be neglected, are almost always initialized as hydrostatic equilibrium torus solutions with simple angular momentum prescriptions \citep{Fishbone1976,Kozlowski1978,Chakrabarti1985,Penna2013}. While such simulations tend to reach a well-defined quasi-steady state in the inner portions of the flow, there is still an open question as to how much this state depends on the fact that the simulations begin with a relatively artificial initial condition.

 In thin accretion disks there is a large timescale separation between the viscous and dynamical times and rapid radiative cooling enables the flow to lose memory of its initial thermodynamic state. This is much less true in RIAFs, so one might expect the dynamics of such flows to retain more memory of how the matter is fed to the vicinity of the black hole from larger radii. There are a number of analytic models for the structure of RIAFs, which differ primarily in the importance of non-radiative energy transport mechanisms, e.g.\ advection, convection/turbulence, and outflows \citep{Narayan1994,Blandford1999,Quataert2000,Narayan2000}. It is unclear which, if any, of these is a correct description of RIAFs or whether different non-radiative energy transport mechanisms are important depending on how matter is fed to the vicinity of the black hole.

 Previous work has shown that the flow structure at small radii in black hole accretion simulations depends on the magnetic field structure threading the initial torus, in particular how much magnetic flux there is at a given radius and/or through the whole torus \citep{Narayan2003,Beckwith2008}. The initial field structure influences both the strength of the resulting jet and whether the dominant angular momentum transport is produced by small-scale turbulent stresses due to the magnetorotational instability or by large-scale magnetic stresses.

One of the computational challenges in studying the connection between the small and large-scale flow structure is that the region in steady state grows outward only sublinearly with simulation time. Here we address this challenge with the brute-force method of long timescale simulations that reach steady state over as extended a radial range as possible. We then attempt to address whether a well-defined self-similar state is reached at small radii independent of initial conditions.

The idea of running GR simulations for long times to achieve steady state over large distances has been pursued before, notably by \citet{Narayan2012}.\footnote{\citet{Stone1999}, \citet{Igumenshchev2000}, and \citet{Yuan2012} studied the radial structure of RIAFs using axisymmetric hydrodynamic simulations with an $\alpha$ viscosity, with outcomes differing with the details of that viscosity. The differences between hydrodynamic and magnetohydrodynamic (MHD) simulations make it difficult to know how to connect the results of those simulations to the MHD problem considered here.} They considered two instances of RIAF flows:\ a disk with standard and normal evolution (SANE), and a magnetically arrested disk (MAD). Their initial conditions differ only in whether the purely poloidal magnetic field has the topology of a single loop in a slice of constant azimuth, conducive to accumulating net vertical magnetic flux and inducing a MAD state, or else has multiple loops and thus stays in the SANE regime. Their SANE simulation reaches steady state out to about $90$ gravitational radii after running for a time $2\times10^5\ GM/c^3$.

Here we focus on the SANE case, primarily because such models are a priori more similar to the self-similar analytic models that still guide much of the intuition in thinking about RIAF structure across many decades in radius. We extend \citet{Narayan2012}'s work by running three variations on the initial conditions to probe how the final state is sensitive to details of the initial state, with the longest simulation run for a time $4.4\times10^5\ GM/c^3$. The numerics and setup of the simulations are detailed in \S\ref{sec:setup}. We present accretion rates and related global quantities in \S\ref{sec:accretion} and radial profiles of quantities in steady state in \S\ref{sec:profiles}. The global inflow and outflow structure is analyzed in \S\ref{sec:structure}, and observational consequences are discussed in \S\ref{sec:observations}. Our conclusions are summarized in \S\ref{sec:conclusion}.

Throughout this work, all length scales will be expressed in implicit units of gravitational radii $GM/c^2$, where $M$ is the black hole mass. Likewise times will be in units of $GM/c^3$ and velocities in units of $c$, with units derived from these following naturally. The magnetic field contains an implicit factor of $1/\sqrt{4\pi}$, so for example the magnetic pressure is given by $\pmag = \eta_{ij} B^i B^j / 2$ if the components $B^i$ are measured in a Minkowski frame comoving with the fluid.

\section{Numerical Setup}
\label{sec:setup}

We perform our simulations with the GRMHD code \athena{} \citep{White2016}. The code uses a second-order van~Leer integrator at a Courant--Friedrichs--Lewy number of $1/4$, where $1/2$ is the maximum stable value for this integrator in three spatial dimensions, along with second-order modified van~Leer spatial reconstruction from \citet{Mignone2014}. Fluxes are calculated via the HLLE Riemann solver.

While less diffusive Riemann solvers such as HLLD exist for relativistic MHD \citep{Mignone2009}, we choose HLLE based on two considerations. First, shocks in HLLD are narrowed to be only one or two cells thick. In a multidimensional finite-volume method, such steep gradients can lead to the conserved variables in a cell entering a physically inadmissible state during a single timestep, manifesting as a variable inversion failure when trying to recover the primitive variables. Given that these simulations need to take tens of millions of timesteps, we want to minimize the chance of catastrophic inversion failures. Second, HLLE is common practice in the GRMHD community. For example, in the recent code comparison of \citet{Porth2019}, which includes \athena{}, the nine codes exclusively employ either HLLE or the more diffusive LLF Riemann solver, yet they are able to reach agreement with sufficient resolution. HLLD would enable us to resolve the same phenomena with fewer cells, but we are willing to use HLLE with higher resolution here in order to achieve robustness over very long integration times.

Spacetime is fixed to be that of a nonspinning black hole, $a = 0$. The simulations employ spherical Kerr--Schild coordinates $(t, r, \theta, \phi)$ with volume element $\sqrt{-g} = r^2 \sin\theta$. Our coordinate system covers the entire sphere, including the appropriate transmissive polar boundary. In the radial direction the cell boundaries are logarithmically spaced from $r = 1.7$ (inside the horizon) to $r = 10^4$. In the poloidal direction we compress the cells near the equatorial plane as in \citet{Gammie2003} (except we adjust cell spacing rather than the metric). The result is that cells are uniformly spaced in a midplane-compressed coordinate $\thetac$ related to the standard $\theta$ via
\begin{equation}
  \theta = \thetac + 0.35 \sin(2 \thetac).
\end{equation}

Due to the particularly long integration times required for these simulations, we additionally use static mesh refinement to keep the part of the grid near the poles at low resolution. The root grid consists of $120 \times 20 \times 20$ cells in the radial, polar, and azimuthal directions respectively. The region $\pi/5 < \abs{\thetac - \pi/2} < 3\pi/10$ is refined by a factor of $2$ in each dimension, and the region $\abs{\thetac - \pi/2} < \pi/5$ is refined by a factor of $4$ relative to the root grid. The cells at the midplane thus have thickness $\Delta\theta \approx 0.0118 \approx \pi/266$.

\begin{figure*}
  \centering
  \includegraphics{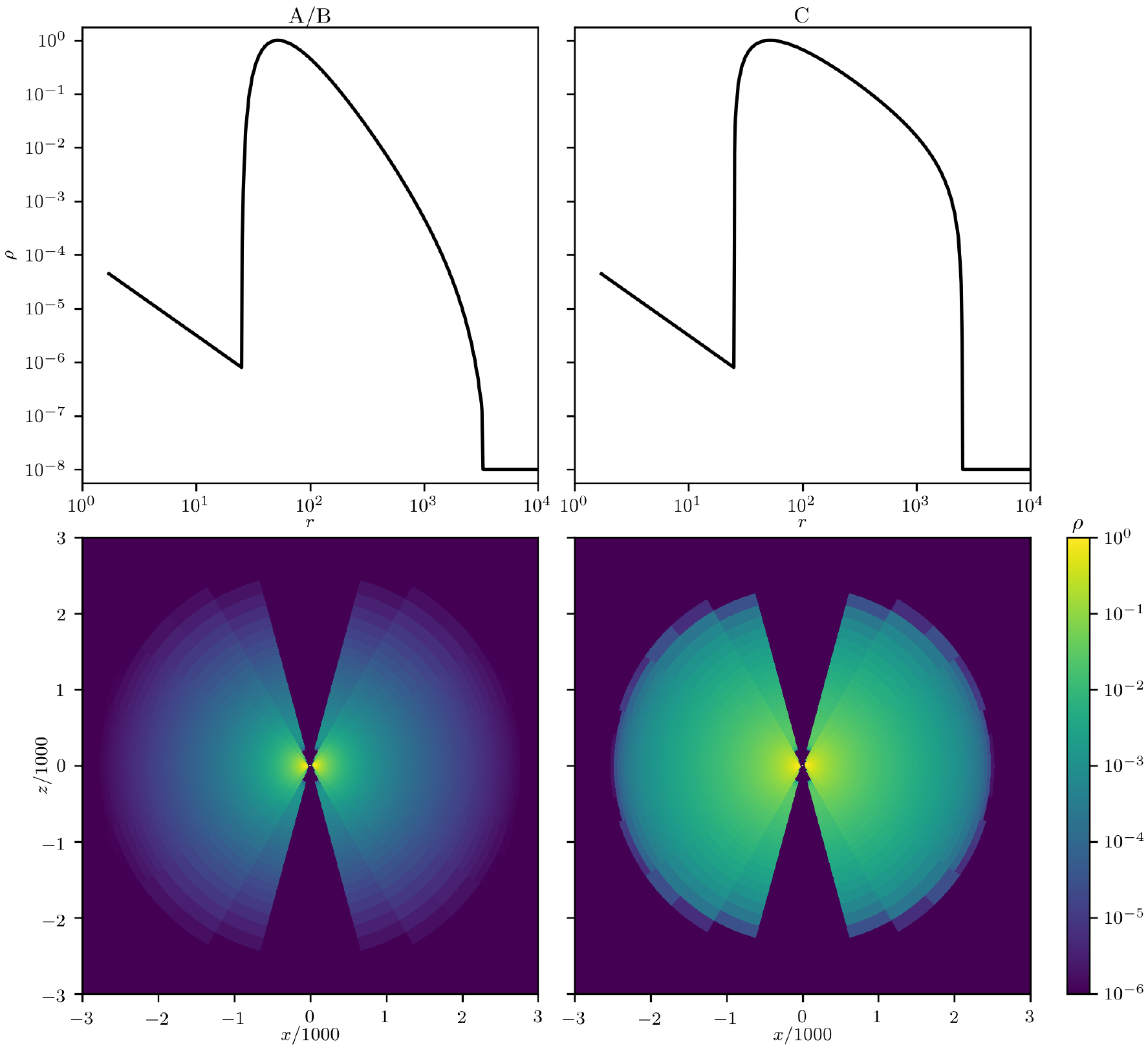}
  \caption{Density $\rho$ for the hydrostatic equilibrium initial conditions. The left panels show the setup for simulations~A and~B, while the right panels show simulation~C. Midplane values as a function of radius are shown on top, with a vertical slice in the $xz$-plane shown on the bottom. The power law floor is a background atmosphere inserted for numerical purposes. \label{fig:initial_density}}
\end{figure*}

The initial conditions are those of the hydrostatic equilibrium solution of \cite{Fishbone1976} with a pressure maximum at $r = 52$ and a peak density of $\rho = 1$. The latter sets an arbitrary scale for the fluid mass, which is taken to be negligible compared to the black hole mass. Simulations~A and~B have an inner edge at $r = 25$ and an adiabatic index of $\Gamma = 4/3$, while simulation~C has an inner edge at $r = 25.1$ and $\Gamma = 5/3$. Indices of $\Gamma = 4/3$ and $\Gamma = 5/3$ are approximations to the thermodynamics appropriate for super-~and sub-Eddington RIAFs, respectively. The slight change of inner edge radii helps to make the extent of the mass distribution of simulation~C similar to that of the other two. The resulting densities are shown in Figure~\ref{fig:initial_density}, with the midplane run of $\rho$ with radius on the top and a vertical slice on the bottom.

\begin{figure}
  \centering
  \includegraphics{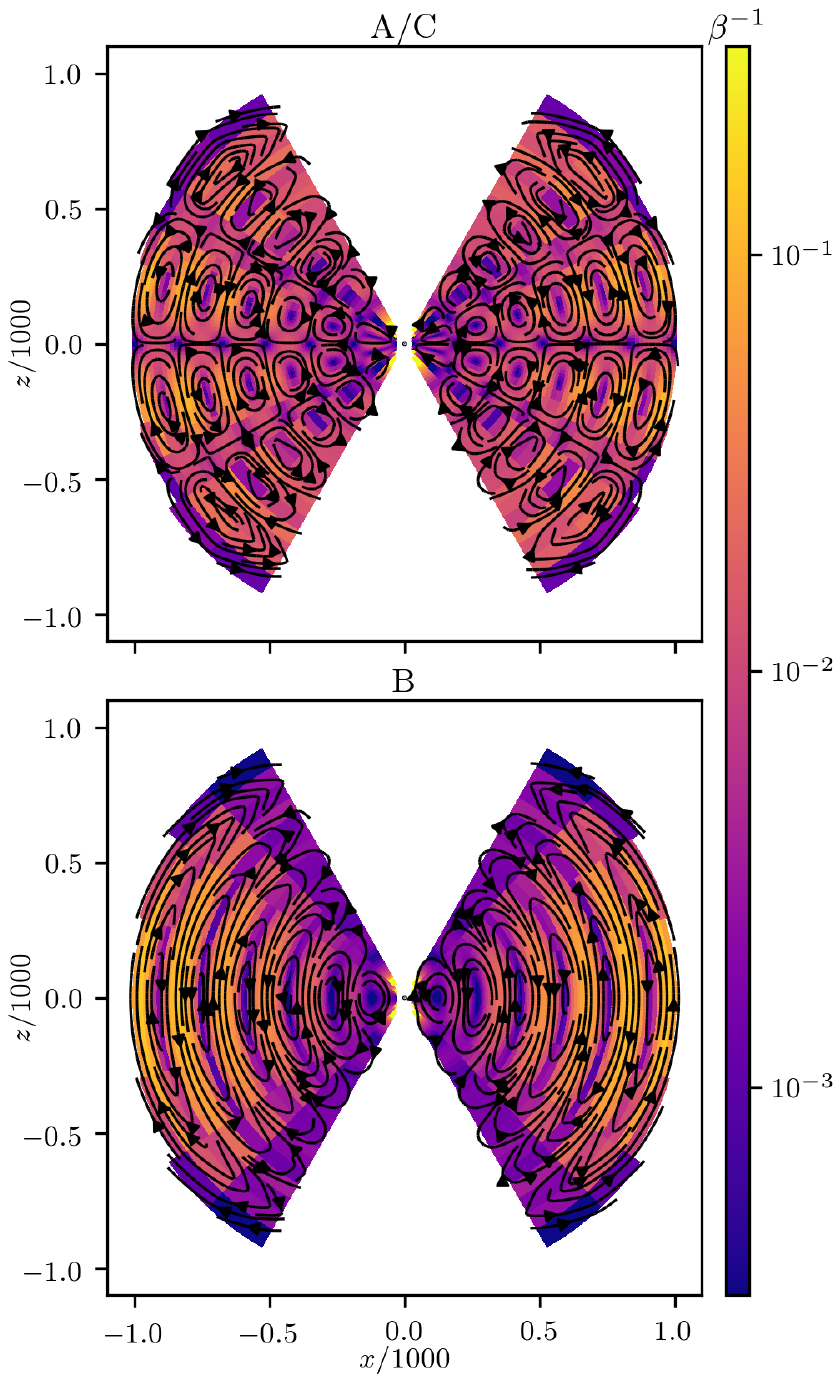}
  \caption{Initial magnetic field configurations for simulations~A and~C (top) and simulation~B (bottom). The streamlines show the poloidal field loops, and the background color is plasma $\beta^{-1}$. While both configurations have no net vertical flux when averaged over sufficiently large volumes, only the one on top has no net vertical flux on any shell of constant radius. Note that the scale of these images is different from that in Figure~\ref{fig:initial_density}; the initial field does not fill the entire torus. \label{fig:initial_magnetization}}
\end{figure}

We add a weak magnetic field to this initial torus. The field is derived from the purely azimuthal vector potential with component
\begin{equation} \label{eq:a_phi}
  A_\phi \propto \paren[\big]{\max(\pgas - \pgasmin, 0)}^{1/2} r^2 \sin(\theta) \sin\paren[\big]{\pi N_r L(r; \rmin, \rmax)} \sin\paren[\big]{\pi N_\theta L(\theta; \thetamin, \thetamax)},
\end{equation}
where $\pgas$ is the gas pressure and $L$ is a linear ramp function:
\begin{equation}
  L(x; \xmin, \xmax) =
  \begin{cases}
    0, & x \leq \xmin; \\
    \dfrac{x-\xmin}{\xmax-\xmin}, & \xmin < x < \xmax; \\
    1, & x \geq \xmax.
  \end{cases}
\end{equation}
The form of this function is chosen to allow for a variable number of counterrotating loops in both directions, and to give roughly constant values for plasma $\beta^{-1} \equiv \pmag / \pgas$, the ratio of magnetic to gas pressure. We choose the parameters $\pgasmin = 10^{-8}$, $\rmin = 30$, $\rmax = 1000$, $\thetamin = \pi/6$, and $\thetamax = 5\pi/6$ in all simulations. For simulations~A and~C we have $N_r = 6$ and $N_\theta = 4$, while in simulation~B we set $N_r = 6$ and $N_\theta = 1$. In all cases the field strength is normalized such that
\begin{equation}
  \frac{\int \beta^{-1} \rho \sqrt{-g} \, \dr\,\dth\,\dph}{\int \rho \sqrt{-g} \, \dr\,\dth\,\dph} = 10^{-2},
\end{equation}
where the integrals exclude the background atmosphere outside the torus. These field configurations are illustrated in Figure~\ref{fig:initial_magnetization}.

Finally, we perturb the initial velocities inside the torus as follows. We consider the normal observer velocity components $u^{i'} \equiv u^i + \beta^i u^0$, where $\beta^i$ is a component of the standard $3{+}1$ shift. We introduce small motions in the poloidal directions according to
\begin{subequations} \label{eq:perturb} \begin{align}
  \Delta u^{1'} & = \Delta_0 \frac{R}{r} u^{3'} \sin(k_R R) \cos(k_z z), \\
  \Delta u^{2'} & = \Delta_0 \frac{z}{r^2} u^{3'} \sin(k_R R) \cos(k_z z),
\end{align} \end{subequations}
where $R = r \sin\theta$ and $z = r \cos\theta$ are standard cylindrical coordinates. We choose $\Delta_0 = 0.03$ and $k_R,k_z = \pi/50$.

Simulation~A is run to $t = 4.4\times10^5$, while the variations $B$ and $C$ are run to $t = 2.2\times10^5$. The cost of these runs is approximately $1.1$ core-hours per simulation time unit (Intel Xeon E5-2670 CPU) or $0.7$ core-hours per time unit (Intel Xeon Platinum 8160 CPU).

\section{Accretion Rates}
\label{sec:accretion}

We follow \citet{Narayan2012} and divide the simulations into logarithmic chunks in time. The eight boundaries of these seven chunks are $3.89\times10^4$, $5.5\times10^4$, $7.78\times10^4$, $1.11\times10^5$, $1.556\times10^5$, $2.2\times10^5$, $3.111\times10^5$, and $4.4\times10^5$. We will use a consistent color scheme to label these chunks throughout all figures.

\begin{figure*}
  \centering
  \includegraphics{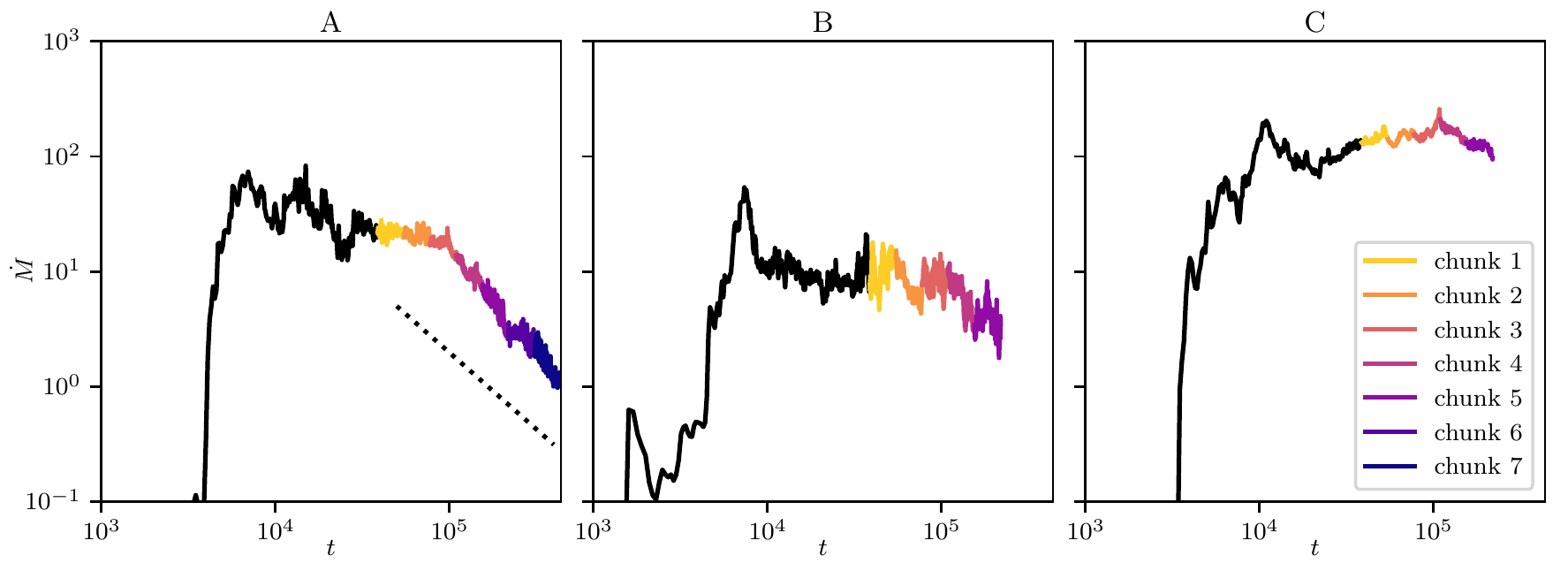}
  \caption{Horizon accretion rates as functions of time. The time chunks used for later analysis are indicated by the colors, where even the earliest chunk begins well after the sudden spike in accretion that is a remnant of the initial conditions. Simulation~C begins with more mass than the other two, and so the torus does not begin to significantly deplete over the course of the simulation, despite the higher accretion rate. The dotted line indicates the slope of a power law going as $t^{-4/3}$. \label{fig:mdot_t}}
\end{figure*}

The mass accretion rate is defined in the usual way, integrating over a spherical shell:
\begin{equation}
  \mdot = -\oint \rho u^1 \sqrt{-g} \, \dth\,\dph.
\end{equation}
The run of the horizon value of $\mdot$ with time is given in Figure~\ref{fig:mdot_t}. As can be seen in the figures, the first time chunks are chosen to be well after the initial transience. Simulations~A and~B begin depleting their mass reservoir, while simulation~C has much more mass and so only shows signs of decreasing $\mdot$ at the very end of the run. By the end of chunk~5 ($t = 2.2\times10^5$), simulations~A, B, and~C have lost $29\%$, $19\%$, and $12\%$ of their initial masses, with $15\%$, $7.0\%$, and $9.5\%$ of the original mass going through the horizon. After simulation~A has run for twice as long, it has lost a total of $61\%$ of its mass, with $17\%$ going into the black hole. Some of this mass loss is due to bulk outflows originating from smaller radii (discussed in \S\ref{sec:structure}), while the rest comes from the outer parts of the initial torus moving outward as they receive angular momentum from the inner parts. In Simulation A, the accretion rate approaches a scaling of $\mdot \sim t^{-4/3}$ at very late times, as shown by the dotted line. This is the expected scaling for a RIAF in the absence of outflows removing mass or angular momentum. Since the simulation has outflows and is probably not fully in the asymptotic regime where $t^{-4/3}$ would be expected, it seems likely that the agreement between the simulation and this analytic expectation is fortuitous.

\begin{figure*}
  \centering
  \includegraphics{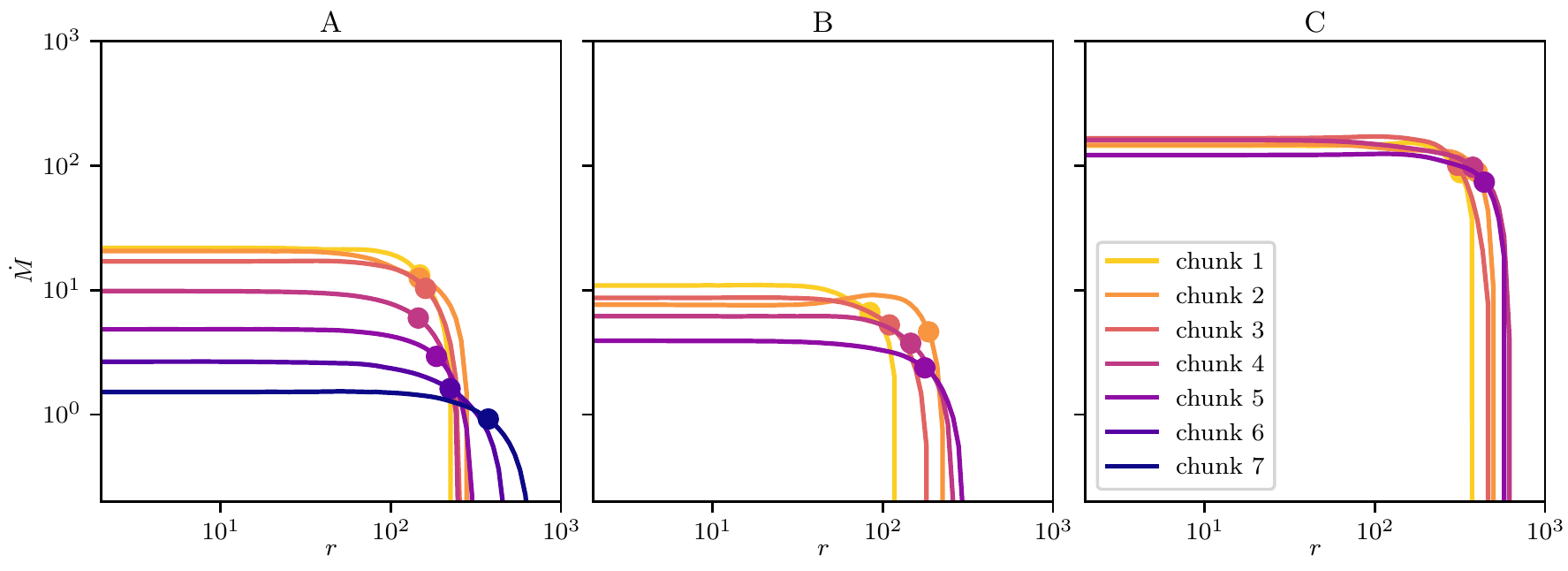}
  \caption{Time-averaged mass accretion rates as functions of radius. Level portions of the curves indicate the establishment of inflow equilibrium. The points denote the outer edges of nominal viscous ranges, which are extending outward as time progresses in all simulations. \label{fig:mdot_r}}
\end{figure*}

We can also look at $\mdot$ as a function of radius. This is done in Figure~\ref{fig:mdot_r}, where the separate lines are time averages over different chunks. In steady state we expect these values to be close to constant out to some radius, and indeed they are even as this constant decreases with time. The flatness of the curves at the inner radii indicates the solution is in a quasi-steady state and that numerical density floors do not play a large role in the inner parts of the simulations.

For later analysis, we use the $\mdot$ vs.\ radius curves to define a ``viscous radius'' inside of which the system has reached inflow equilibrium. We choose the point where $\mdot$ drops to $\ee^{-1/2}$ times its horizon value. These radii are delineated by the points in Figure~\ref{fig:mdot_r}. The range over which equilibrium is established does indeed move outward in time, extending to $r = 370$ in chunk~7 of simulation~A and to $r = 170$ and $r = 440$ in chunk~5 of simulations~B and~C. For a constant $h/r$ and constant $\alpha$ model we would expect the viscous radius to move outward in time as $t^{2/3}$. The numerical evolution is generally slower than this, as we discuss further below in \S\ref{sec:profiles}.

\begin{figure*}
  \centering
  \includegraphics{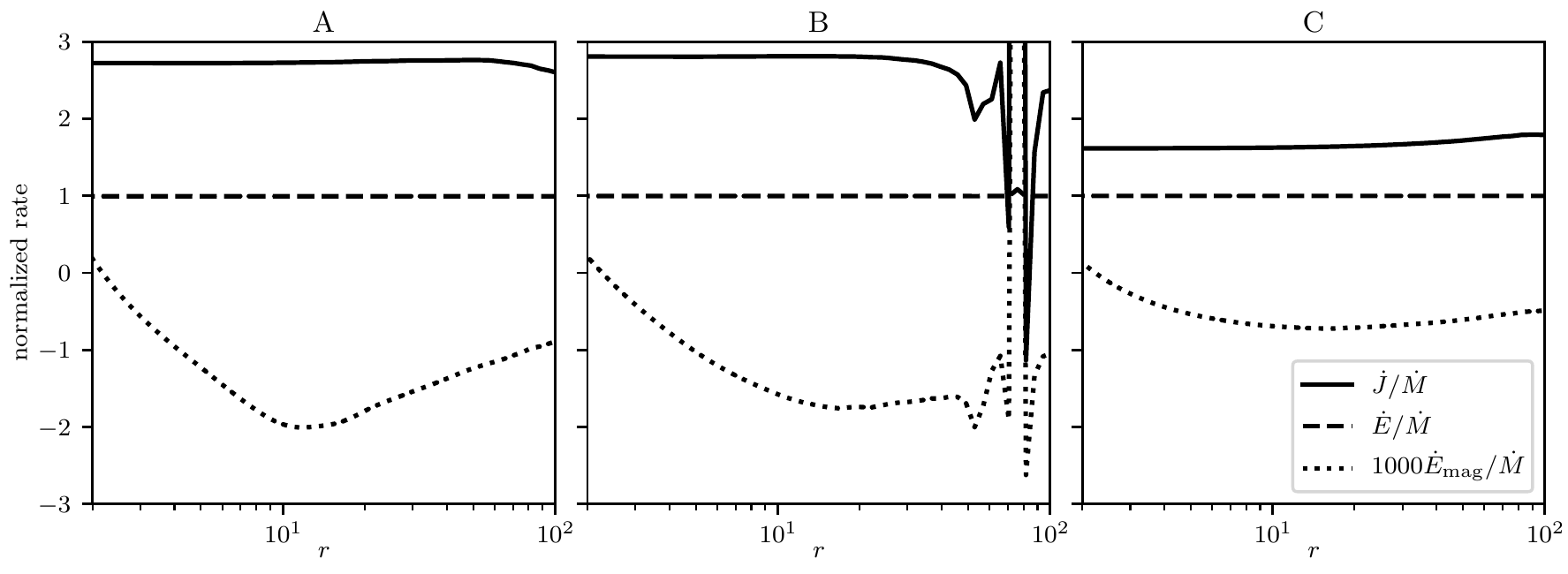}
  \caption{Time-averaged energy and angular momentum accretion rates as functions of radius. In all cases the rates are normalized by the horizon value of $\mdot$ before being averaged in time over the fifth time chunk ($1.556\times10^5 \leq t < 2.2\times10^5$). Both $\jdot$ and $\edot$ should be constant in steady state, as they are, with $\edot/\mdot \approx 1$ indicating rest mass advection dominates the energy budget. Electromagnetic energy outflow is a small fraction of the total (a consequence in part of our use of non-spinning black holes). The negative sign of $\edotmag$ corresponds to outflow of energy as measured at infinity. \label{fig:edot_r}}
\end{figure*}

The energy accretion rate is very similar to the total mass accretion rate. Define
\begin{subequations} \begin{align}
  \edot & = \oint \tensor{T}{^1_0} \sqrt{-g} \, \dth\,\dph, \\
  \edotmag & = \oint \tensor{(T_\mathrm{mag})}{^1_0} \sqrt{-g} \, \dth\,\dph,
\end{align} \end{subequations}
where the stress-energy tensor is given by
\begin{subequations} \begin{align}
  T^{\mu\nu} & = T_\mathrm{gas}^{\mu\nu} + T_\mathrm{mag}^{\mu\nu}, \\
  T_\mathrm{gas}^{\mu\nu} & = \paren[\bigg]{\rho + \frac{\Gamma}{\Gamma-1} \pgas} u^\mu u^\nu + \pgas g^{\mu\nu}, \\
  T_\mathrm{mag}^{\mu\nu} & = 2 \pmag u^\mu u^\nu + \pmag g^{\mu\nu} - b^\mu b^\nu, \\
  \pmag & = b_\lambda b^\lambda, \\
  b^0 & = u_i B^i, \\
  b^i & = \frac{1}{u^0} \paren[\big]{B^i + b^0 u^i}.
\end{align} \end{subequations}
Runs of $\edot$ and $\edotmag$ with radius, averaged in time over chunk~5, are shown in Figure~\ref{fig:edot_r}. The other time chunks show essentially the same behavior. The overall energy accretion rate tracks mass very closely. This is to be expected, since rest mass density $\rho$ dominates over gas pressure $\pgas$ as well as the contribution of the electromagnetic field. That is, we are mostly measuring rest mass being advected into the black hole. The electromagnetic contribution to the flux of energy, shown by the dotted line, is negative, indicating an outflow. However the magnitude of this component is only one part in a thousand of the total. This would be significantly larger for a spinning black hole.

Just as with mass and energy we monitor the flux of angular momentum, defining
\begin{equation}
  \jdot = -\oint \tensor{T}{^1_3} \sqrt{-g} \, \dth\,\dph.
\end{equation}
This is also shown in Figure~\ref{fig:edot_r}. Again we see the radially constant plateaus indicating approximate steady state, though the deviations begin at slightly smaller radii than for $\mdot$. The steady-state value of $\jdot / \mdot$ does not change with magnetic field topology (cf.\ simulations~A and~B), but does change in the case of a different $\Gamma$ (simulation~C).

\begin{figure}
  \centering
  \includegraphics{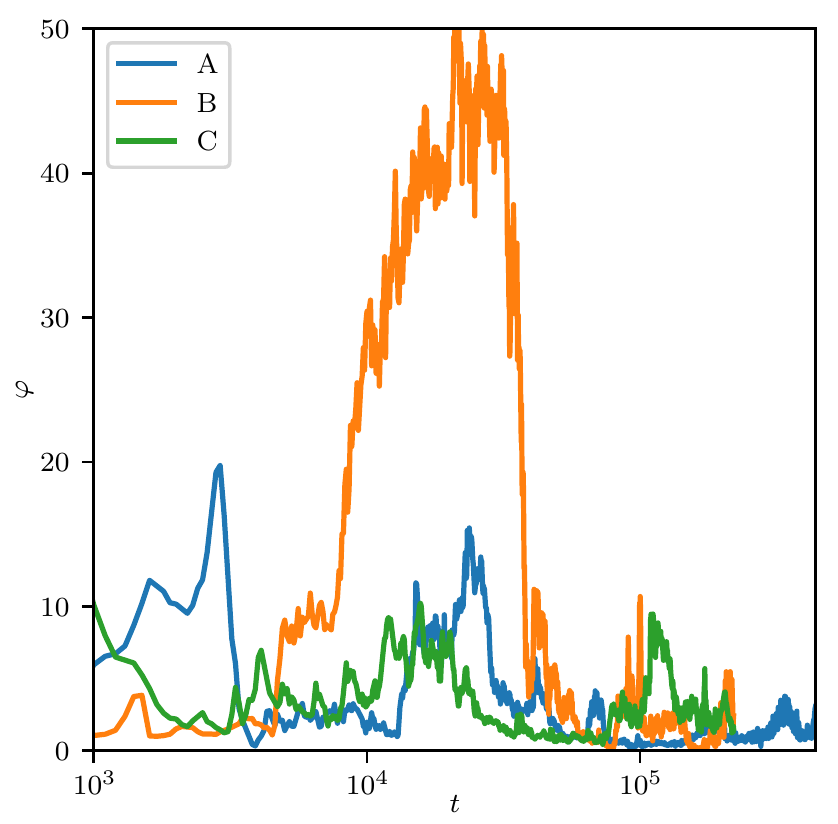}
  \caption{Horizon-penetrating magnetic fluxes as functions of time. Simulation~B does not stay in a MAD state, but it does build up more coherent flux than simulations~A and~C. This is not unexpected given the initial conditions (Figure~\ref{fig:initial_magnetization}), where each radius in simulation~B has net vertical flux while those in simulations~A and C do not. \label{fig:phi_t}}
\end{figure}

We additionally calculate the horizon-penetrating flux
\begin{equation}
  \varphi = \frac{\sqrt{4\pi}}{2 \mdot^{1/2}} \oint\limits_{r=\rhor} \abs{B^1} \sqrt{-g} \, \dth\,\dph.
\end{equation}
This is the same quantity as $\phi_\mathrm{BH}$ defined in \citet{Tchekhovskoy2011} and \citet{Narayan2012}, where our explicit factor of $\sqrt{4\pi}$ is implicit in their units, except here we use the instantaneous $\mdot$ not convolved with any smoothing kernel in time. The run of $\varphi$ with time in the two simulations is shown in Figure~\ref{fig:phi_t}. Simulations~A and~C, with their small initial field loops (see Figure~\ref{fig:initial_magnetization}), stay very much in the SANE regime, with $\varphi \lesssim 10$ at late times. For comparison, the MAD simulations in \citet{Tchekhovskoy2011} have $\varphi \approx 47$. Simulation~B goes through a burst of higher flux, though it does not stay high as in a true MAD disk. The behavior of simulation~B matches that of the SANE simulation in \citet{Narayan2012}, which uses a similar initial field topology.

\section{Radial Profiles in the Disk}
\label{sec:profiles}

We now turn to characterizing the properties of the disks, starting with profiles of disk properties as functions of radius.

\begin{figure*}
  \centering
  \includegraphics{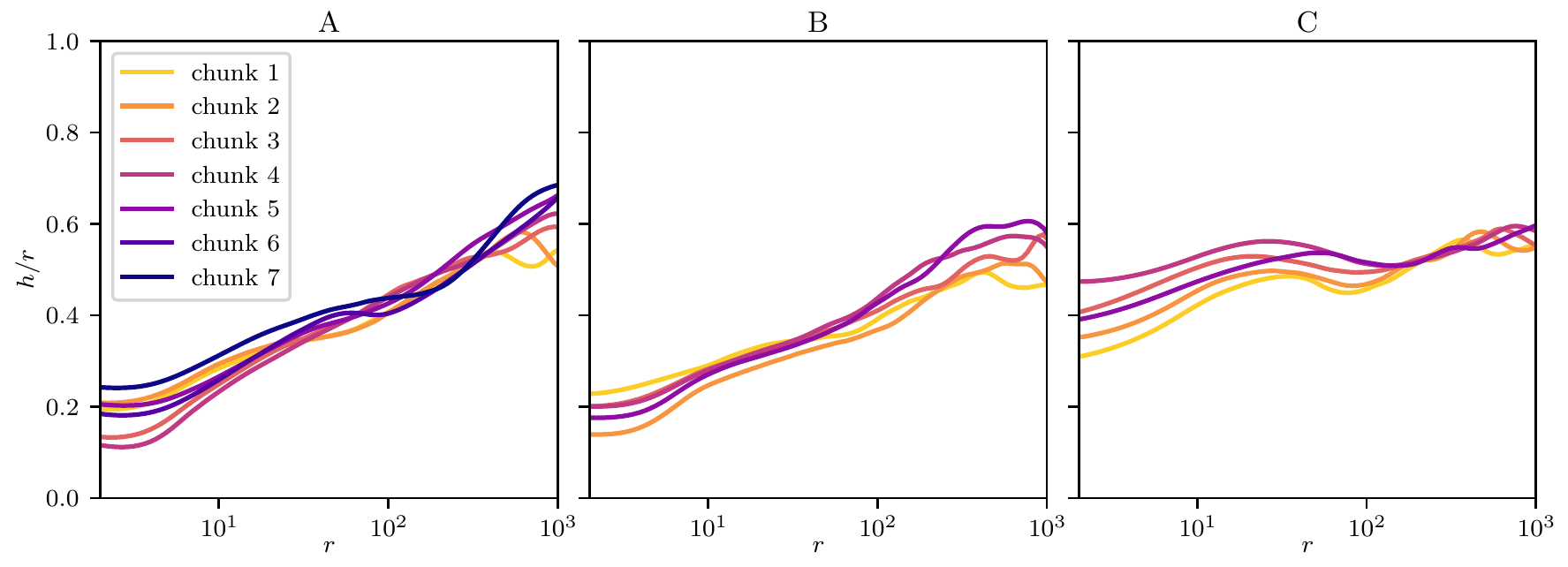}
  \caption{Time-averaged scale heights as functions of radius. The disks have roughly the same thicknesses in all three cases, though simulation~C is somewhat more spherical overall (see also Figure~\ref{fig:field}). The scale heights do not change much over time. \label{fig:hr_r}}
\end{figure*}

We define the spherical scale height at a radius to be
\begin{equation}
  \frac{h}{r} = \tan\paren[\bigg]{\frac{\oint \abs{\theta - \pi/2} \rho \sqrt{-g} \, \dth\,\dph}{\oint \rho \sqrt{-g} \, \dth\,\dph}},
\end{equation}
where the integrals are taken over spheres at that radius. The run of this quantity with radius, averaged within time chunks, is shown in Figure~\ref{fig:hr_r}. In all simulations, this scale height stays between approximately $0.1$ and $0.7$, increasing toward larger radii and changing little over time. Simulation~C, with $\Gamma = 5/3$, stays somewhat thicker at very small radii compared to the other cases.

With scale height in hand, we can define averages of quantities over the disk proper, here taken to be the region within one scale height of the midplane. The average density we define to be
\begin{equation}
  \ave{\rho} = \frac{\int_\mathrm{disk} \rho \sqrt{-g} \, \dth\,\dph}{\int_\mathrm{disk} \sqrt{-g} \, \dth\,\dph}.
\end{equation}
Note that a more appropriate volume element might use the determinant of the metric restricted to $2{+}1$ hypersurfaces of constant $r$, or even $2{+}0$ surfaces of constant $t$ and $r$. However all these choices differ by factors that depend only on $r$ in the case of nonspinning black holes, and so they all lead to equivalent definitions of $\ave{\rho}$.

\begin{figure*}
  \centering
  \includegraphics{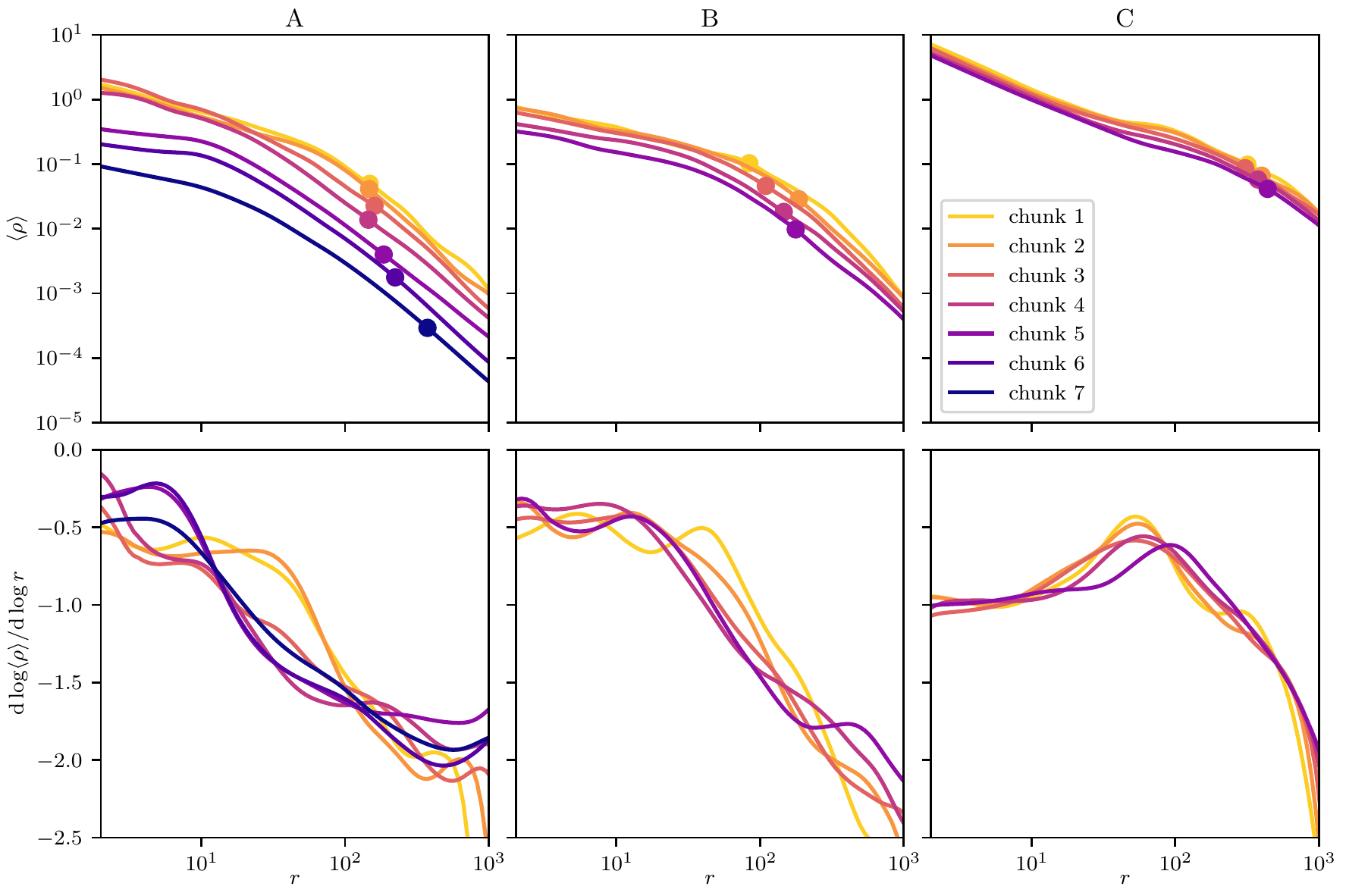}
  \caption{Density profiles in radius, averaged over the disk and averaged over time chunks. The profiles themselves are displayed in the top panels, with their power-law slopes shown below. The points are at the same viscous radii as in Figure~\ref{fig:mdot_r}. Simulations~A and~B show shallower slopes inside $r \approx 10$ compared to further out, while simulation~C has an intermediate slope at small radii and a shallow plateau at intermediate radii. There is no convergence to a well-defined density profile at intermediate radii (between the ISCO and viscous radii), as one might expect if the solutions were approaching a self-similar RIAF solution. \label{fig:rho_r}}
\end{figure*}

The profiles of density are shown in the upper panels of Figure~\ref{fig:rho_r}, where the points indicate inflow equilibrium as defined in \S\ref{sec:accretion}. The profiles generally approach power laws in radius, especially at very late times in simulation~A. The very inner regions of simulations~A and~B, inside $r \approx 10$, have a shallower power-law slope when compared with the outer part of the viscous range. This is likely produced by the rapid acceleration needed close to the horizon:\ conservation of mass implies that the density profile flattens as the velocity accelerates. Somewhat surprisingly, simulation~C does not display this knee, though it does have a slight persistent kink between $r = 50$ and $r = 100$.

The lower panels of Figure~\ref{fig:rho_r} show the power-law slopes of the density profiles. Specifically, for an abscissa $r_0$ the ordinate is the slope of a linear regression to the set of points $(\log r, \log\ave{\rho})$ for which $r$ is within a factor of approximately $2$ of $r_0$. The inner regions of simulations~A and~B have slopes of very roughly $-1/2$. The range over which this holds does not increase after a point, even at very late times as in simulation~A. The middle of the inflow equilibrium range in simulation~A eventually tends toward a slope of approximately $-3/2$ at very late times. Simulation~B does not reach sufficiently late times and does not show signs of converging to a constant slope. Simulation~C, on the other hand, has a slope of between $-1/2$ and $-1$ over all radii in equilibrium. Analytic models of RIAFs predict power-law density profiles far from the horizon ranging from $r^{-3/2}$ \citep{Narayan1994} to $r^{-1/2}$ \citep{Narayan2000,Quataert2000}, and everything in between \citep{Blandford1999}. One might expect that there would be a well-defined self-similar state of the accretion flow that obtains at radii $r \gg 1$, where the impact of the black hole horizon on the flow structure is no longer significant. The density profiles in Figure \ref{fig:rho_r} are not consistent with this ansatz. Instead, the profiles depend on the thermodynamics, parameterized by adiabatic index here, and initial magnetic field structure.

We can average other quantities in a density-weighted way. For any quantity $q$ other than $\rho$, define
\begin{equation}
  \ave{q} = \frac{\int_\mathrm{disk} q \rho \sqrt{-g} \, \dth\,\dph}{\int_\mathrm{disk} \rho \sqrt{-g} \, \dth\,\dph}.
\end{equation}
Although not shown explicitly here, all three accretion flows are slightly sub-Keplerian, as can be seen by examining $\ave{\Omega}$ for Boyer--Lindquist angular velocity $\Omega = u_\mathrm{BL}^3 / u_\mathrm{BL}^0$. The average infall velocity $-\ave{v^r}$, where $v^r = u^1 / u^0$, is well below the free-fall value.

\begin{figure*}
  \centering
  \includegraphics{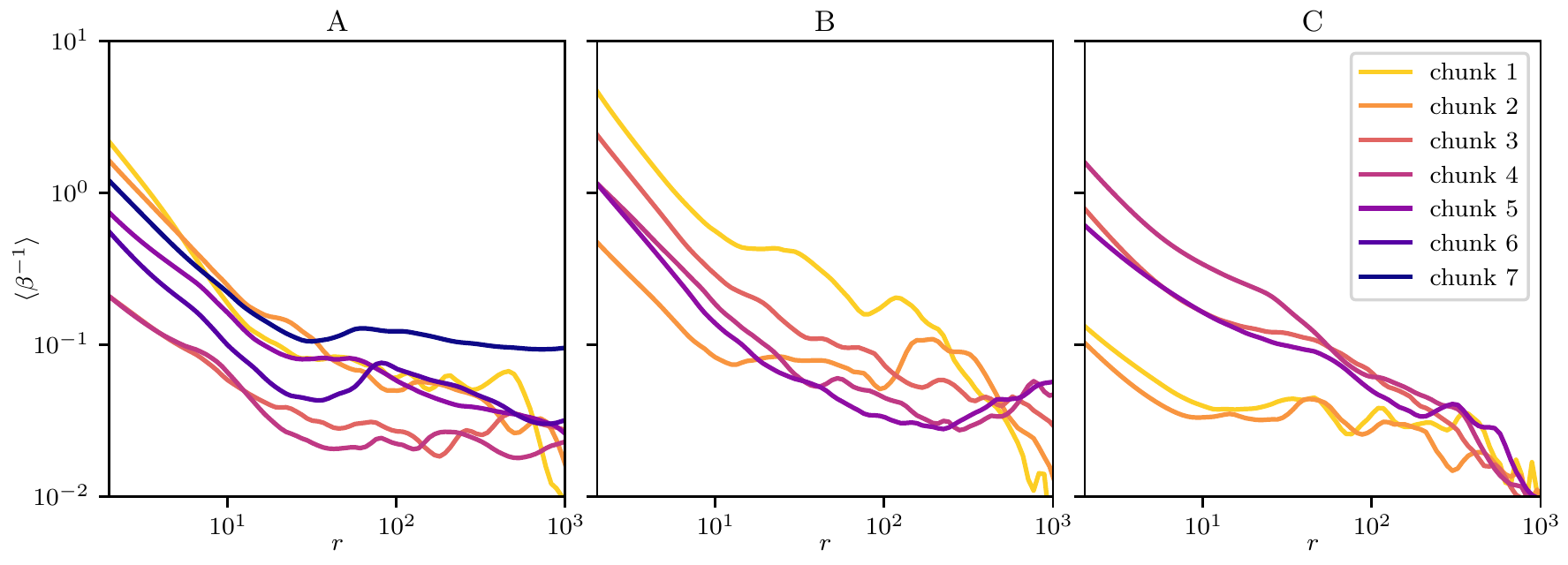}
  \caption{Profiles of magnetic-to-gas pressure ratios in radius, averaged in a density-weighted way over the disk and then averaged over time chunks. In all cases the magnetization steeply declines with radius in the inner parts of the disk and tends to level off at large radii. Note that the trends with time are not necessarily monotonic. \label{fig:beta_inv_r}}
\end{figure*}

The averages of magnetization $\ave{\beta^{-1}}$ are given in Figure~\ref{fig:beta_inv_r}. Magnetization decreases with increasing radius out to $r \approx 25$ in all simulations, but it remains relatively flat beyond that in cases~A and~B, even when equilibrium is established much further out. The levels of these flat portions can change in time, however, first decreasing and then increasing at late times (simulation~A). We note that simulations~A and~B, whose only difference is the initial field topology, reach the same profile of $\ave{\beta^{-1}}$ by the fourth time chunk, indicating that the saturated field strength may not strongly depend on details of initial conditions. In simulation~C, the profile of $\ave{\beta^{-1}}$ levels off near where $\ave{\rho}$ does, and then proceeds to again decrease with increasing radius beyond that.

The increased magnetization at small radii influences the rates at which our viscous radii move outward. Given the scaling $\tvisc \sim r^{3/2} (h/r)^{-2}$, and given the observed thicknesses $h/r \sim r^{1/4}$ (Figure~\ref{fig:hr_r}), we might expect equilibrium to be established out to a radius $\rvisc \sim t$ by time $t$. However the radii marked in Figure~\ref{fig:mdot_r} do not advance in time this rapidly. For example, chunks~1 and~7 are separated by a factor of $8$ in time, while equilibrium is only established $2.5$ times further out in the latter compared to the former for simulation~A. This can be explained by noting we also expect $\tvisc \sim \alpha^{-1}$, and $\alpha \sim \beta^{-1}$. Thus $\rvisc \sim \beta^{-2/3} t$, and so equilibrium is established much more rapidly at small radii with strong magnetization than at large radii.

\section{Global Structure of the Accretion Flow}
\label{sec:structure}

\begin{figure*}
  \centering
  \includegraphics{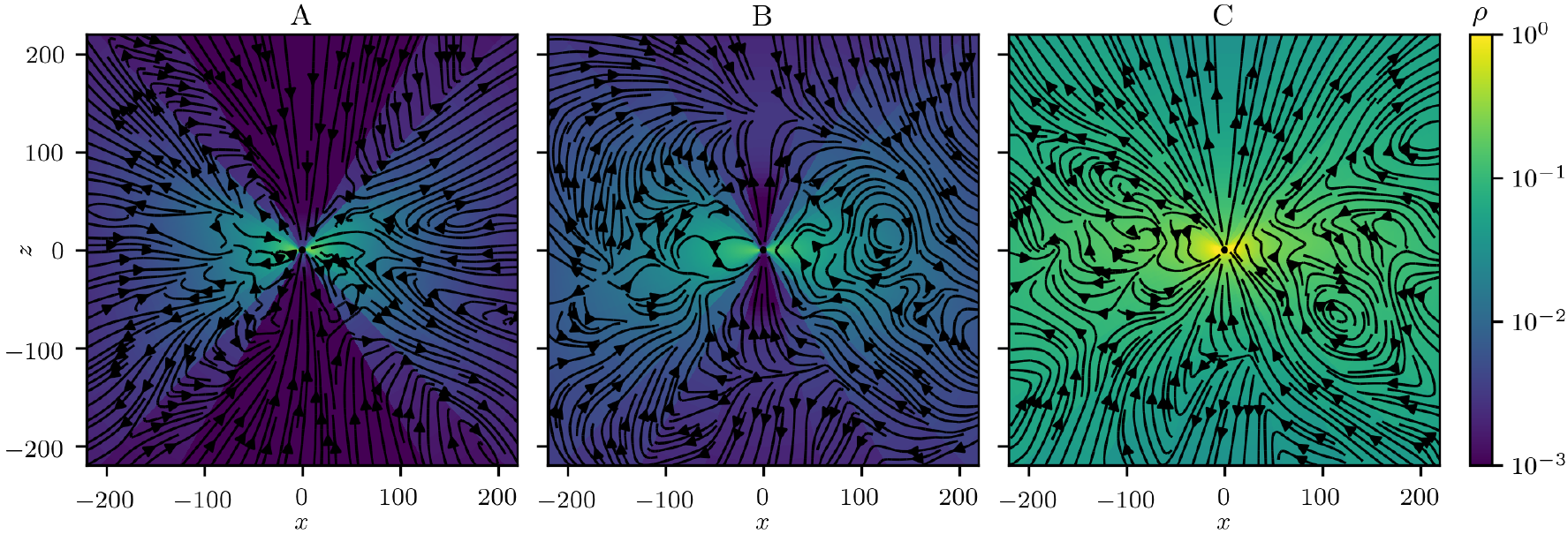}
  \caption{Average late-time (chunk~5, $1.556\times10^5 \leq t < 2.2\times10^5$) poloidal magnetic field, with background color indicating average density. Simulation~A displays more coherent radial field in the disk than the other two. Additionally, the symmetric/antisymmetric character of $B^z$ across the midplane $z = 0$ follows from the character of the initial fields, shown in Figure~\ref{fig:initial_magnetization}. The density structure in simulation~C is quite spherically symmetric compared to the equatorial disk and evacuated polar region seen in simulation~A. \label{fig:field}}
\end{figure*}

Going beyond radial profiles, we show the late-time magnetic field in the poloidal plane in Figure~\ref{fig:field}, superimposed on the time-averaged density. This magnetic field is obtained by averaging $B^1$ and $B^2$ in azimuth and averaging in time over the last common time chunk (5) available for all simulations. It shows what steady-state structure arises from the initial fields in Figure~\ref{fig:initial_magnetization}.

We can immediately see that simulation~A has a more ordered, radial field near the midplane when compared to the other two cases, despite simulation~C having the same initial field topology. This can be explained by the velocity structure of the flows, discussed below. Simulation~A also shows near cancellation in vertical fluxes at fixed radius; i.e.\ $B^z$ is antisymmetric across $z = 0$. Simulation~B, on the other hand, shows a field topology with $B^z$ symmetric across the midplane. While the net flux integrated over a large radial extent may vanish, there is generally net flux at any given radius. These symmetries are also found in the initial conditions, indicating that even at these late times the initial field's imprint on the field topology remains. It is also striking how spherical the density distribution is in Simulation C:\ there is only a factor of few variation in density from equator to pole. By contrast simulations~A and~B show the more ``standard'' disc structure of a dense midplane and a low-density polar region.

The net accretion rates shown in Figures~\ref{fig:mdot_t} and~\ref{fig:mdot_r} do not fully describe the radial motion of matter, since there can be outflows partially negating mass inflow elsewhere. While this will certainly happen over small time and length scales due to turbulence, coherent outflows may persist even when averaged over turbulence.

\begin{figure*}
  \centering
  \includegraphics{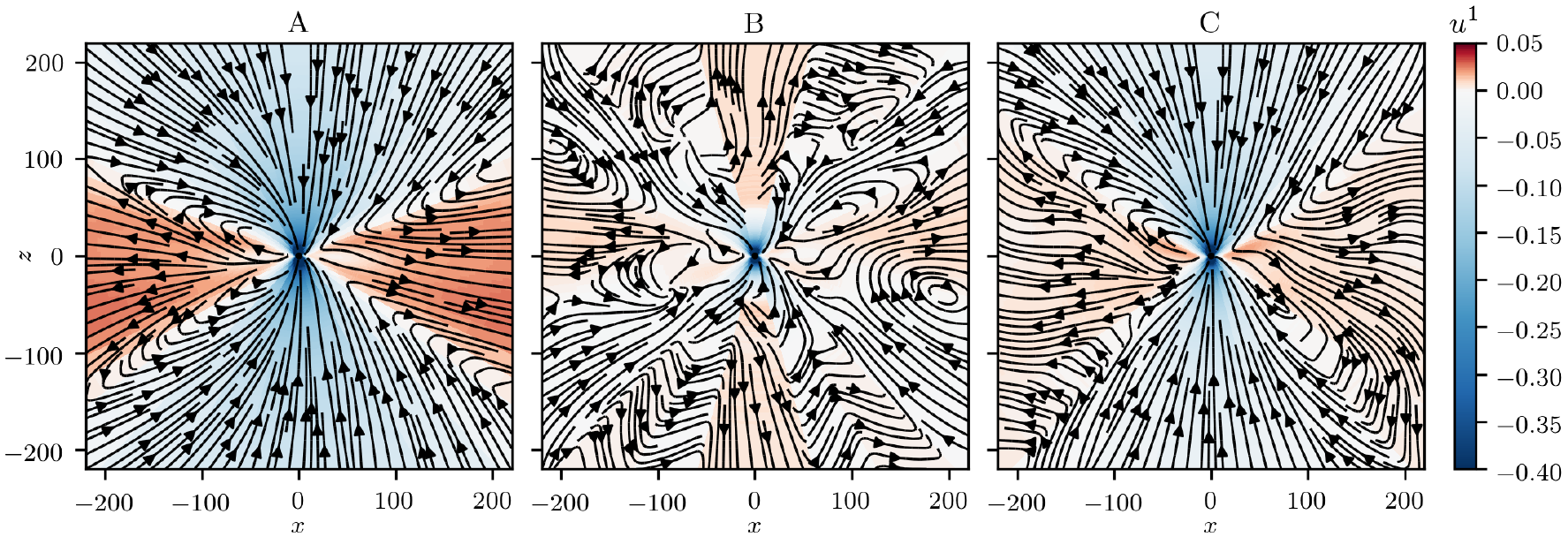}
  \caption{Average late-time (chunk~5, $1.556\times10^5 \leq t < 2.2\times10^5$) poloidal velocity field in the simulations. The background colors highlight the radial velocity, with blue indicating inflow. Simulations~A and~C display polar inflows with outflows in the disk, while simulation~B has polar outflows and at least some disk inflow. \label{fig:velocity}}
\end{figure*}

Indeed, we do observe large-scale outflows in our simulations, particularly with more ordered magnetic fields. Figure~\ref{fig:velocity} shows the velocity field for each simulation in the poloidal plane, averaged in azimuth with density weighting and again averaged over the last common time chunk. Simulation~B shows an accretion disk with a low radial velocity, together with fast, low-density, polar outflows. This is similar to the now-canonical assumed structure of RIAFs from earlier simulations (e.g., \citealt{Hawley2002}). Simulations~A and~C, however, show the reverse. In these cases matter is falling in through the poles, with outflow occurring in the disk. This disk outflow pattern is particularly strong in simulation~A.

\begin{figure*}
  \centering
  \includegraphics{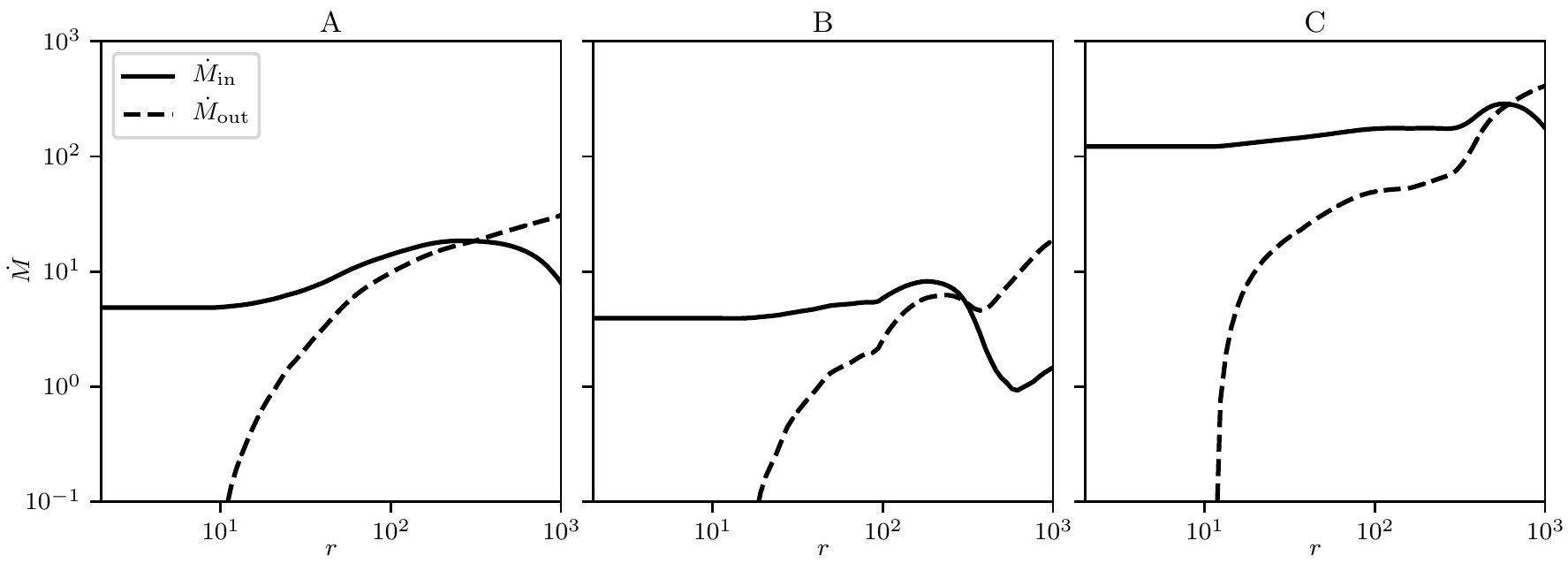}
  \caption{Bulk ingoing and outgoing mass fluxes at late times (chunk~5, $1.556\times10^5 \leq t < 2.2\times10^5$) in the simulations. Over the viscous range, inflow is relatively constant, while outflow is less in magnitude and decreases inward. Over the factor of $\sim 300$ in radius that our solutions are in viscous equilibrium, the total outflow rate is only comparable to the inflow rate. There is no bulk outflow inside $r \approx 10$, indicating that this is inside the stagnation radius of any polar outflows (simulation~B, see Figure~\ref{fig:velocity}). Note that the ingoing and outgoing mass fluxes necessarily combine to the net inflow rates shown Figure~\ref{fig:mdot_r}. \label{fig:mdot_in_out}}
\end{figure*}

We can further decompose $\mdot$ into ingoing and outgoing components based on radial mass fluxes that have been averaged in time and azimuth:
\begin{subequations} \begin{align}
  \mdotin & = -2\pi \int_0^\pi \min(\ave{\rho u^1}, 0) \sqrt{-g} \, \dth, \\
  \mdotout & = 2\pi \int_0^\pi \max(\ave{\rho u^1}, 0) \sqrt{-g} \, \dth.
\end{align} \end{subequations}
The runs of these quantities with radius are given in Figure~\ref{fig:mdot_in_out}. We know $\mdotin - \mdotout$ must be constant in radius from Figure~\ref{fig:mdot_r}, and indeed $\mdotin$ is close to constant, with $\mdotout$ being a small perturbation over most of the viscous range. The fact that $\mdotout$ becomes negligible inside $r \approx 10$ in simulations~A and~C tells us that the disk outflows we observe (see Figure~\ref{fig:velocity}) do not extend all the way to the horizon. In simulation~B, $\mdotout$ also drops off at small radii, matching how the polar outflow shown in Figure~\ref{fig:velocity} reaches a stagnation surface at $r \approx 25$. Our results on the disk outflows are broadly consistent with \citet{Narayan2012}, who also find that outflows were at most an order unity perturbation to the accretion rate for the radial range they simulate.

What is the physical origin of the outflows seen in Figure \ref{fig:mdot_in_out} and the flow structure seen in Figure \ref{fig:velocity}? Two possible hydrodynamic mechanisms for large-scale flows of this kind are convection and meridional circulation. We examine these quantitatively in Appendix~\ref{sec:conv_circ} and find that they are not very compelling in explaining our numerical results.

The fact that both simulations~A and~C display polar inflows and disk outflows despite having different values of $\Gamma$ and different initial tori indicates this phenomenon is not limited to a small set of hydrodynamic initial conditions. Simulations~A and~B, meanwhile, have identical hydrodynamic initial conditions yet vastly different steady-state velocity structures. This points to the original magnetic field topology as being important in determining the inflow and outflow structure (as was emphasized by, e.g., \citealt{Beckwith2008}).

As discussed in \S\ref{sec:setup}, the initial field in simulation~B consists of elongated loops in the poloidal plane (see Figure~\ref{fig:initial_magnetization}). While there is no net vertical flux when averaged over sufficiently large radial extents, there is net flux at any given radius. Thus accretion brings a sequence of alternating net fluxes to the black hole over time (Figure~\ref{fig:phi_t}), causing bursts in accretion rate (Figure~\ref{fig:mdot_t}). While the black hole has no spin and thus there are no Blandford--Znajek jets \citep{Blandford1977}, the net flux in the disc may be enough to drive a polar outflow via the Blandford--Payne mechanism \citep{Blandford1982} or related MHD processes. This can prevent any loosely bound polar inflow coming in from the disk at larger radii from reaching the black hole.

Simulations~A and~C, on the other hand, have no net flux at any radius. It is more difficult for the magnetic field in these cases to launch polar outflows, and so material can fall to the black hole near the polar axes. With enough polar inflow, the liberated gravitational potential energy can help to unbind material, driving it outward in the equatorial plane.

\begin{figure*}
  \centering
  \includegraphics{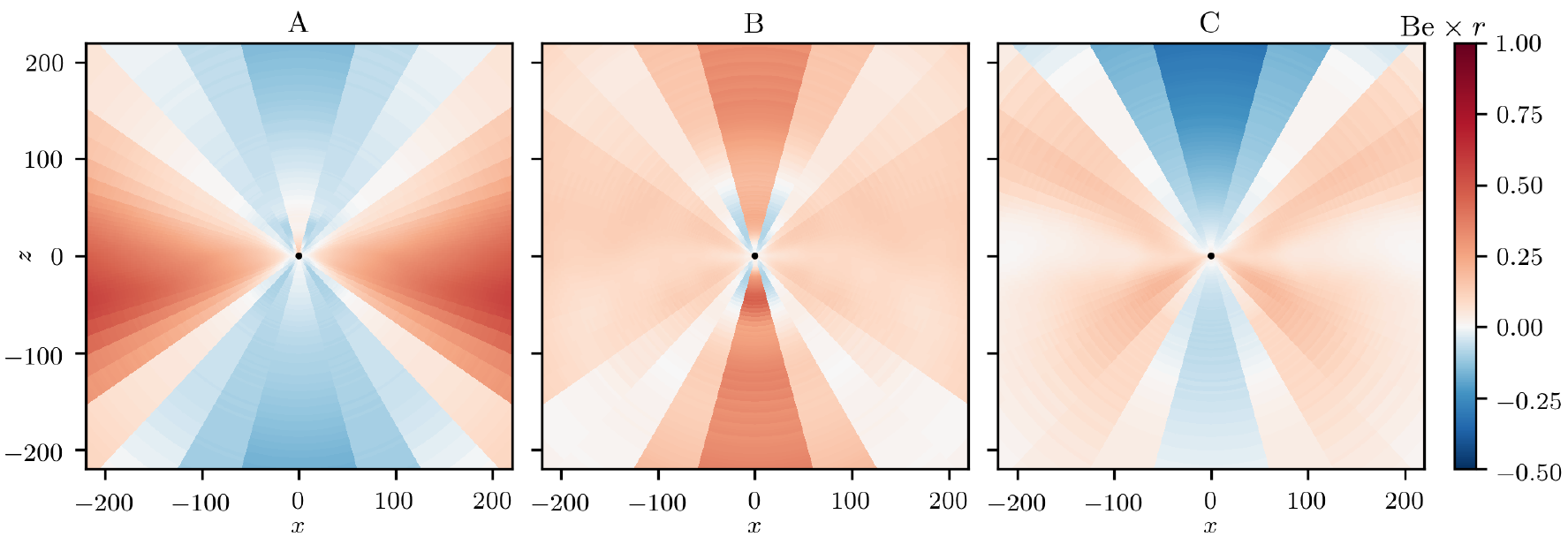}
  \caption{Average Bernoulli parameter (see equation~\eqref{eq:bernoulli}) in the simulations, calculated in the fifth time chunk ($1.556\times10^5 \leq t < 2.2\times10^5$). Only in simulation~B is the polar region unbound. The disks have largely become unbound, despite the fact that the initial conditions are bound everywhere with negative Bernoulli parameter. \label{fig:bernoulli}}
\end{figure*}

We quantify the extent to which material is bound with the Bernoulli parameter, which we define as
\begin{subequations} \label{eq:bernoulli} \begin{align}
  \be & \equiv -\frac{\ave{w u_0}}{\ave{\rho}} - 1, \\
  w & \equiv \rho + \frac{\Gamma}{\Gamma-1} \pgas + 2 \pmag,
\end{align} \end{subequations}
with averages being taken in azimuth and time. This is similar to equation~\extref{11} of \citet{Narayan2012}. Plots of $\be$ for the last common time chunk of each simulation are given in Figure~\ref{fig:bernoulli}. Note $\be < 0$ throughout the initial tori.

We see that most of the disks in all three cases have become unbound. At the same time, the polar regions of simulations~A and~C are bound, while they are very much unbound in simulation B. The initial magnetic field topology thus has a strong effect on the overall energetics and dynamical structure of RIAFs. This occurs primarily, we believe, via the energy redistribution and outflows associated with magnetic stresses, not just the angular momentum transport they produce.

\begin{figure*}
  \centering
  \includegraphics{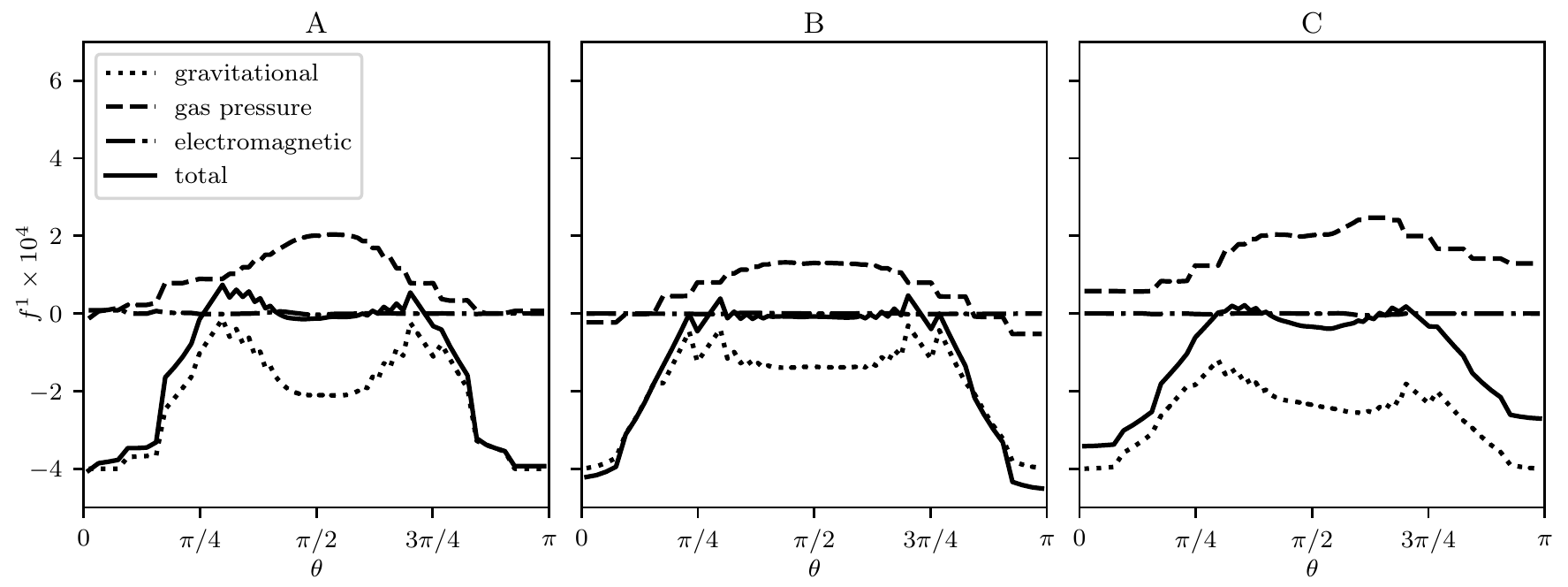}
  \caption{Forces acting in the radial direction (see equation~\eqref{eq:force_radial}) for a slice at $r = 50$, calculated in the fifth time chunk ($1.556\times10^5 \leq t < 2.2\times10^5$). The balance of forces is similar in all cases, with a disk supported by rotation and gas pressure and with inward accelerations at high latitudes. \label{fig:force_radial}}
\end{figure*}

Evidence for this comes from the fact that over a large range of radii all three simulations have similar radial accelerations, vanishing near the midplane and directed inward at high latitudes. The presence of a narrow outflow along the polar axis, for instance, is not due to consistent, outward, electromagnetic acceleration along the axis, but rather to the combined effect of accelerations at other locations. We quantify these accelerations by writing the inhomogeneous geodesic equation for a perfect fluid as
\begin{subequations} \label{eq:force_radial} \begin{align}
  \frac{\dd u^\mu}{\dtau} & = f_\mathrm{grav}^\mu + f_\mathrm{gas}^\mu + f_\mathrm{EM}^\mu, \\
  f_\mathrm{grav}^\mu & = -\tensor*{\Gamma}{^\mu_{\alpha\beta}} \tensor{u}{^\alpha} \tensor{u}{^\beta}, \\
  f_\mathrm{gas}^\mu & = -\frac{1}{w} \tensor{P}{^{\mu\alpha}} \tensor{\nabla}{_\alpha} \pgas, \\
  f_\mathrm{EM}^\mu & = -\frac{1}{w} \tensor{P}{^{\mu\alpha}} \paren[\big]{\tensor{\nabla}{_\alpha} \pmag - \tensor{\nabla}{_\beta} (\tensor{b}{_\alpha} \tensor{b}{^\beta})},
\end{align} \end{subequations}
where $P^{\mu\alpha} = g^{\mu\alpha} + u^\mu u^\alpha$ is the projection operator relative to the fluid velocity. Figure~\ref{fig:force_radial} shows the radial components of these three forces along an arc of constant radius $r = 50$, computed from the fluid state obtained by averaging in time and azimuth over chunk~5. The gravitational (including centrifugal) force is closely balanced by the gas pressure gradient; the magnetic pressure gradient and Maxwell stress are both small. In simulation~A but not in simulation~C, there may be some ongoing outward radial acceleration near the upper and lower disk surfaces. In these locations $f_\mathrm{grav}^1$ approaches $0$, indicating a large centrifugal force nearly balancing gravitational attraction even without pressure support.

Appendix~\ref{sec:small} shows the steady-state behavior of two additional runs, paralleling simulations~A and~B in how their initial fields differ but using a smaller initial torus. The dichotomy between polar inflows and outflows persists, with inflows occurring where there is no local net vertical flux in the initial conditions. These smaller tori, which are closer in size to those most often employed in the literature, are more tightly bound, and so even the case with polar inflow does not develop an equatorial outflow. The latter result highlights that conclusions drawn from small torus simulations might be inaccurate in some cases.

\section{Observational Consequences}
\label{sec:observations}

Given the length of time we have run these simulations, we have a large amount of data on their time variability. In particular, for simulation~A we have high-cadence samples of the horizon value of $\mdot$ sampled with a time interval $\Delta t = 1$. For a Sgr~A*-like mass of $4.3\times10^6\ \msun$, this corresponds to a sampling period of $21\ \second$.

\begin{figure}
  \centering
  \includegraphics{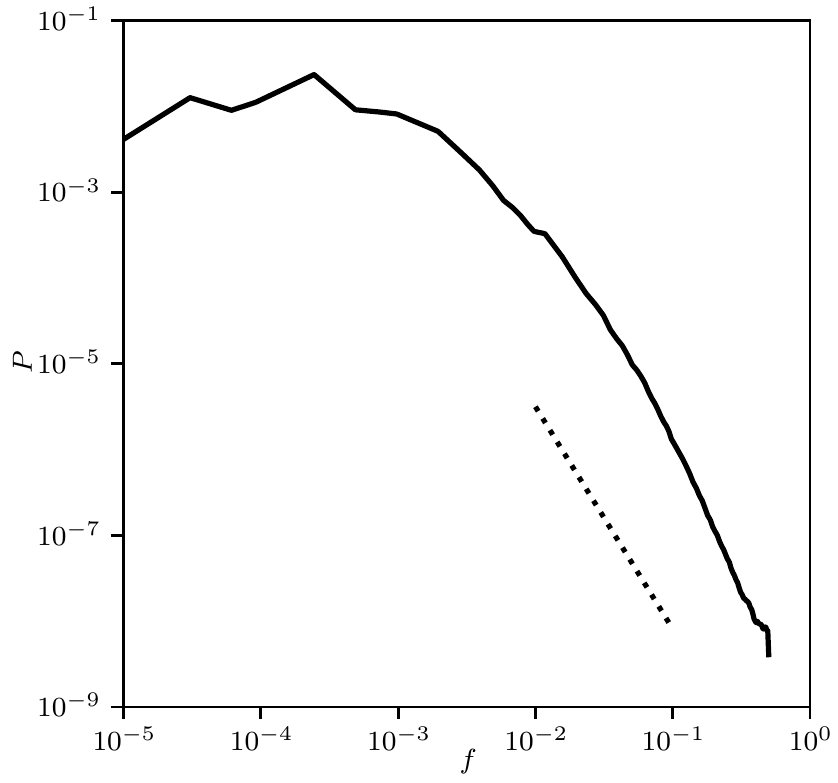}
  \caption{Power spectrum for time series of $\mdot$ at the horizon in simulation~A. The data is taken from Chunks~6--7, covering $2.2\times10^5 \leq t \leq 4.4\times10^5$. The dotted line shows a slope of $-2.59$, which fits the spectrum over the indicated range. \label{fig:variability}}
\end{figure}

We construct the power spectrum of the time series using Welch's method on overlapping segments of time durations ranging from $2^7$ (used to calculate a smooth spectrum at high frequencies) to $2^{15}$ (used at low frequencies), applying a Hann window to each segment. An overall linear trend in the quantities $\log t$ and $\log\mdot$ is subtracted before calculating each spectrum. We do this for the union of chunks~6 and~7. The results are shown in Figure~\ref{fig:variability}. Here we refer to $P = f P_f$ as the power, where $P_f$ is the power per unit frequency.

The spectrum has a slope $\dd\log P / \dd\log f \sim -2.59$ over the range $10^{-2} \leq f \leq 10^{-1}$, somewhat steeper than ideal red noise. The slope is $-2.38$ if we analyze chunks~1--5 instead of 6 and 7. A turnover to a flat spectrum is apparent at approximately $f = 10^{-3}$, especially in simulation~A. Simulations~B and~C have similar spectra at low frequencies, both in amplitude and slope. However, their $\mdot$ values are only sampled every $100$ time units, and so we cannot measure their spectra at frequencies above $5\times10^{-3}$.

Numerical models of RIAFs have been widely used to predict event-horizon-scale emission and images. Such models form the basis for interpreting Event Horizon Telescope images of M87 \citep{EHT2019a,EHT2019b,EHT2019c} and Sgr~A*. Since our simulations provide three different realizations of the near-horizon structure of RIAFs, it is valuable to assess whether their predictions for the near-horizon emission differ. To do so, we employ the code \ibothros{} \citep{Noble2007} to generate images of synchrotron emission at $230\ \ghz$ in post processing. We take $644$ snapshots from chunk~5, separated by $\Delta t = 100$. For each one, we assume thermal synchrotron radiation from the electrons, with a simulation-to-electron temperature ratio of $4$. The density scale of the simulation is adjusted to match the observed flux of $2.4\ \jy$ of Sgr~A* \citep{Doeleman2008} assuming we are viewing the disk edge-on.

\begin{figure*}
  \centering
  \includegraphics{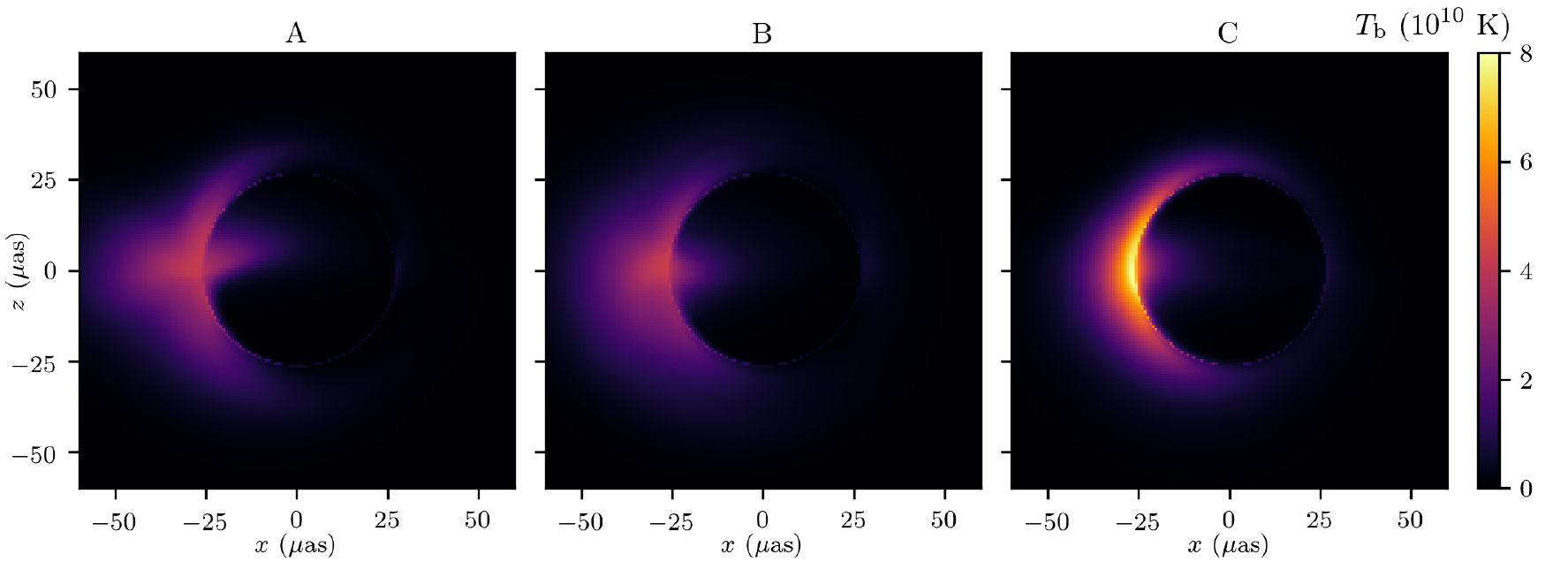}
  \caption{Images of the average $230\ \ghz$ thermal synchrotron emission from the three simulations, modeled using parameters appropriate for an edge-on view of Sgr~A*. Simulations~A and B have images more dominated by the equatorial disk while simulation~C shows a more prominent photon ring. This is a consequence of the differences in the scale heights and density profiles of these simulations (Figures~\ref{fig:hr_r} and~\ref{fig:rho_r}). All three images have the same total flux. \label{fig:images}}
\end{figure*}

Figure~\ref{fig:images} displays the average image resulting from stacking the $644$ frames for each model. Simulation~A shows the most extended disk component outside the typical bright ring around the circular dark region. This equatorial component is most compact in simulation~C, brightening the ring but not increasing its width on the approaching side of the disk. These differences can be understood via the average density and temperature structures in the flows. Case~C has the steepest density profile in the inner region (Figure~\ref{fig:rho_r}), as well as higher temperatures (as can be seen in the larger scale height, Figure~\ref{fig:hr_r}). Thus the emission we see will be dominated by a small region near the photon orbit, slightly spread out along the ring but not radially. In contrast, case~B and especially case~A have emission from a wider range of radii near the disk midplane.

\section{Conclusion}
\label{sec:conclusion}

We have run and analyzed three long timescale simulations of RIAFs around non-spinning black holes. Building on \citet{Narayan2012} (see also related hydrodynamic work by \citealt{Yuan2012}) our goal has been to understand whether a well-defined self-similar structure of the accretion flow develops that is independent of the simulation initial conditions, and whether such a state is described by existing analytic models of RIAFs. All of our simulations are ``SANE,'' i.e.\ they do not have significant net magnetic flux over an appreciable range of radii. We run a fiducial simulation (A) with small alternating field loops (see Figure~\ref{fig:initial_magnetization}) and adiabatic index $\Gamma = 4/3$ to $t = 4.4\times10^5$. Variations with either larger loops (B) or $\Gamma = 5/3$ (C) are run to $t = 2.2\times10^5$. Note that $\Gamma = 4/3$ and $\Gamma = 5/3$ are rough approximations to super-~and sub-Eddington RIAFs, respectively. The simulations reach inflow equilibrium, defined here as where $\mdot$ falls to $\ee^{-1/2}$ of its horizon value, past $r = 370$, $r = 170$, and $r = 440$, respectively. Energy and angular momentum fluxes are as constant with radius as mass fluxes over much of this range. We note that many of our principal conclusions are not unique to these long-timescale runs. On the contrary, Appendix~\ref{sec:small} shows that they hold even for the smaller tori typically used in GRMHD simulations.

One of our principal results is that the dynamical structure of RIAFs is sensitive to initial conditions, at least over the time-~and length-scales that we can afford to simulate. There is no evidence for a unique flow structure that loses memory of the initial conditions. This is not necessarily surprising given the lack of timescale separation among thermal, viscous, and dynamical times in RIAFs. Moreover, the absence of radiative cooling implies that the dynamics in RIAFs is particularly sensitive to modest changes in flow energetics (produced, e.g., by differences in magnetic field structure).

In more detail, Figure~\ref{fig:rho_r} shows that the $\Gamma = 4/3$ simulations maintain shallow slopes in density of $\dd\log\ave{\rho} / \dd\log r \approx -1/2$ in the innermost regions, with slopes at least as steep as approximately $-3/2$ in the outer regions that are still in steady state. In the $\Gamma = 5/3$ case, we see slopes closer to $-1$ in the inner regions and $-1/2$ in the outer regions. In a given simulation, we find that there is no clear convergence of $\dd\log\ave{\rho} / \dd\log r $ to a particular value in the steady state flow exterior to the inner $10\text{--}30\ GM/c^2$, as might be expected if there were a unique self-similar solution for the structure of RIAFs. By contrast, such a well-defined radial profile is seen in hydrodynamic $\alpha$ models of RIAFs (e.g., \citealt{Stone1999,Yuan2012}). It is unclear if this difference between the hydrodynamic and MHD models reflects the very different angular momentum and energy transport physics in the two different simulations or if our simulations still do not have sufficient dynamic range to reach a quasi self-similar state. We suspect the former.

The larger magnetic field loops in simulation~B relative to simulation~A (see Figure~\ref{fig:initial_magnetization}) result in buildup of coherent magnetic flux at the horizon. While this simulation does not stay in a MAD state, it shows signs of alternating between ``semi-MAD'' states with oppositely directed fluxes. This does not have a noticeable effect on radial profiles of scale height, density, or magnetization, but it can affect the velocity structure of the accretion flow. In particular, the coherent vertical flux is enough to drive polar outflows in simulation~B, whereas both simulations~A and~C have polar inflow of material (Figure~\ref{fig:velocity}). The absence of polar outflows in Simulation~C is particularly striking as the density profile becomes nearly spherical with only a factor of few density contrast between the equator and poles (Figure~\ref{fig:field}).

The polar inflows we find are not readily explained by models for convective stability or meridional circulation (see Appendix~\ref{sec:conv_circ}). However even a small amount of inflow can release enough energy to unbind significant portions of the disk, which for RIAFs is at best marginally bound. This is likely why the flow structure is sensitive to the magnetic field initial conditions, since the latter influences the angular momentum and energy transport at essentially all times in the simulation. Overall, we find that unbound outflows change the horizon-scale accretion rate by only a factor of order unity relative to the accretion rate at $r \sim 300$ (see Figure~\ref{fig:mdot_in_out}), consistent with \citet{Narayan2012}.

The sensitivity of the flow structure and dynamics in our simulations to modest changes in initial conditions (all within the context of highly artificial initial conditions) suggests that there is likely to be significant variation in the properties of RIAFs in nature depending on exactly how matter is supplied to the vicinity of the black hole from larger radii. Understanding this better in future work would be very valuable and would impact problems as diverse as accretion in the Galactic Center from stellar winds (e.g., \citealt{Ressler2018}) to the growth of supermassive black holes by highly super-Eddington accretion (e.g., \citealt{Begelman2017}). We note that our calculations assume a nonspinning black hole, and future work should explore whether spin exacerbates or mitigates the sensitivity to how matter is supplied.

Numerical models of RIAFs have been widely used to predict event-horizon-scale emission, variability and images, particularly in the context of the well-studied systems M87 \citep[e.g.][]{Dexter2012,Moscibrodzka2016,Moscibrodzka2017,Ryan2018,Chael2019} and Sgr~A* \citep[e.g.][]{Moscibrodzka2014,Ressler2017,Chael2018}. As a result, we have made preliminary attempts to understand the observational implications of the diversity of flow structure and dynamics we find in our RIAF simulations. Towards this end, we calculate the power spectrum of $\mdot$ in time for simulation~A, using high-cadence samples at time intervals of $1\ GM/c^3$ ($21\ \second$ for $M = 4.3\times10^6\ \msun$), shown in Figure~\ref{fig:variability}. The spectral power goes as $P \sim f^{-2.59}$ at high frequencies. Our data shows a turnover toward white noise ($f^0$) at low frequencies; such a turnover has been measured in the submillimeter emission of Sgr~A*, occurring at a timescale of $\tau \approx 8\ \hour$ \citep{Dexter2014}. This timescale corresponds to a frequency in our geometric units of $f = 7\times10^{-4}$ (for reference, this is the reciprocal of the orbital period at $r = 36$), which is consistent with where we see the turnover in $\mdot$ power.

Motivated by the Event Horizon Telescope, we also calculate millimeter images of synchrotron emission for parameters appropriate for Sgr~A* (see Figure~\ref{fig:images}). All of our different realizations of RIAFs produce qualitatively similar images, dominated by a Doppler boosted region along with a faint ring (Figure~\ref{fig:images}). There are, however, interesting differences in the size of the emitting region and the prominence of the equatorial accretion disk, which is much less dominant in Simulation~C ($\Gamma = 5/3$) relative to Simulations~A and B ($\Gamma = 4/3$).

\acknowledgments

We thank all of the members of the Horizon Collaboration (http://horizon.astro.illinois.edu) for useful discussions. We thank {J.~M.} Stone for valuable comments on the manuscript. This research was supported in part by the National Science Foundation under grants NSF~PHY-1748958, NSF~AST~13-33612, and NSF~AST~1715054; by \emph{Chandra} theory grant TM7-18006X from the Smithsonian Institution; and by a Simons Investigator award from the Simons Foundation (EQ). This work used the Extreme Science and Engineering Discovery Environment (XSEDE) Stampede2 at the Texas Advanced Computing Center through allocation AST170012, as well as the Savio computational cluster resource provided by the Berkeley Research Computing program at the University of California, Berkeley.

\software{\athena{} \citep{White2016}, \ibothros{} \citep{Noble2007}}

\appendix

\section{Analysis of Convection and Meridional Circulation}
\label{sec:conv_circ}

\begin{figure*}
  \centering
  \includegraphics{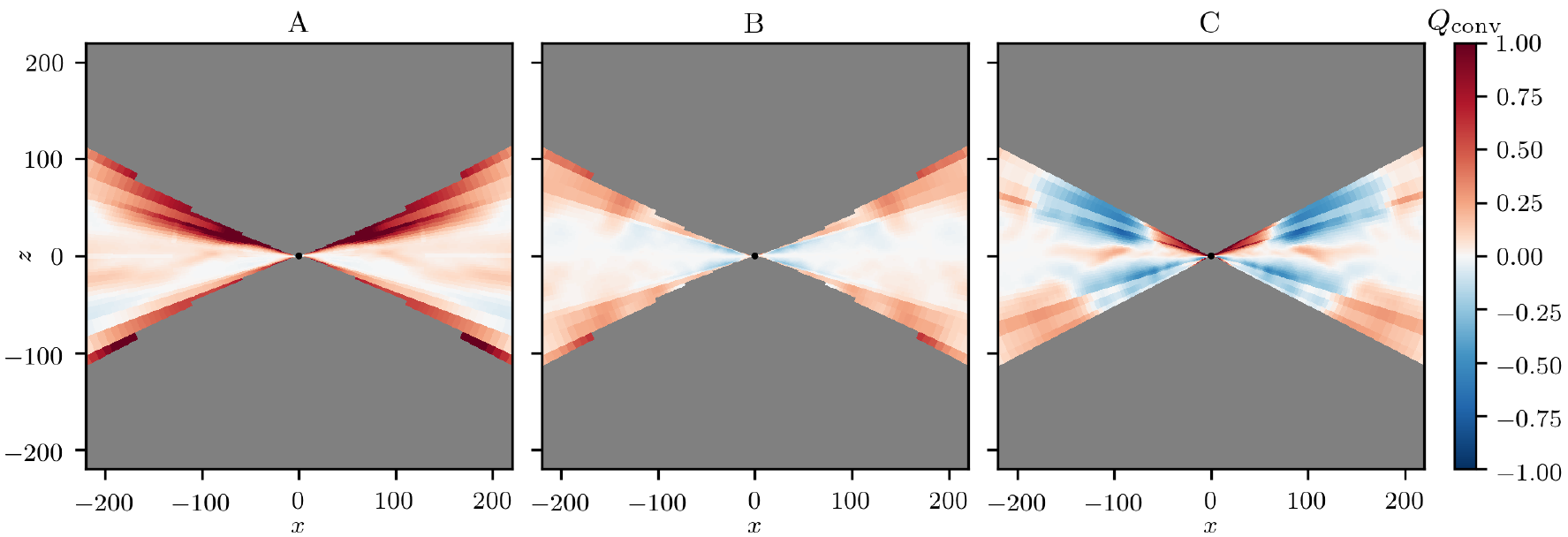}
  \caption{Convective stability measure $\qconv$ \citep{Quataert2000} within one scale height of the midplane for the fifth time chunk ($1.556\times10^5 \leq t < 2.2\times10^5$) of each simulation. Convection is expected for positive values. While simulations~A and~C show outflows in the disk, they are no less stable than simulation~B. \label{fig:stability}}
\end{figure*}

One possible cause of bulk outflows in disks is a stationary convection pattern. As described in \citet{Quataert2000} we can check for convective stability in the disk by examining the gradients of entropy $s \equiv \log(\pgas \rho^{-\Gamma})$ and azimuthal velocity $v^\phi \equiv r \sin(\theta) u^3 / u^0$. We construct the quantity
\begin{equation}
  \qconv = r^6 (C^2 - 4 A B),
\end{equation}
where $A$, $B$, and $C$ are defined by \extref{A2--A4} of \citeauthor{Quataert2000}, using density-weighted averages of $s$ and $v^\phi$ in time and azimuth and using the last time chunk $1.556\times10^5 \leq t < 2.2\times10^5$ available to all simulations. Stability requires $\qconv < 0$. The values for $\qconv$ within one scale height of the midplane are shown in Figure~\ref{fig:stability}. The two simulations with disk outflows, A and C, are slightly unstable against convection in the midplane, in contrast to simulation~B, which is marginal. However simulation~C is quite stable at slightly higher latitudes, where there is still a strong outflow. Thus convection does not seem to fully explain the different bulk patterns shown in Figure~\ref{fig:velocity}.

\begin{figure}
  \centering
  \includegraphics{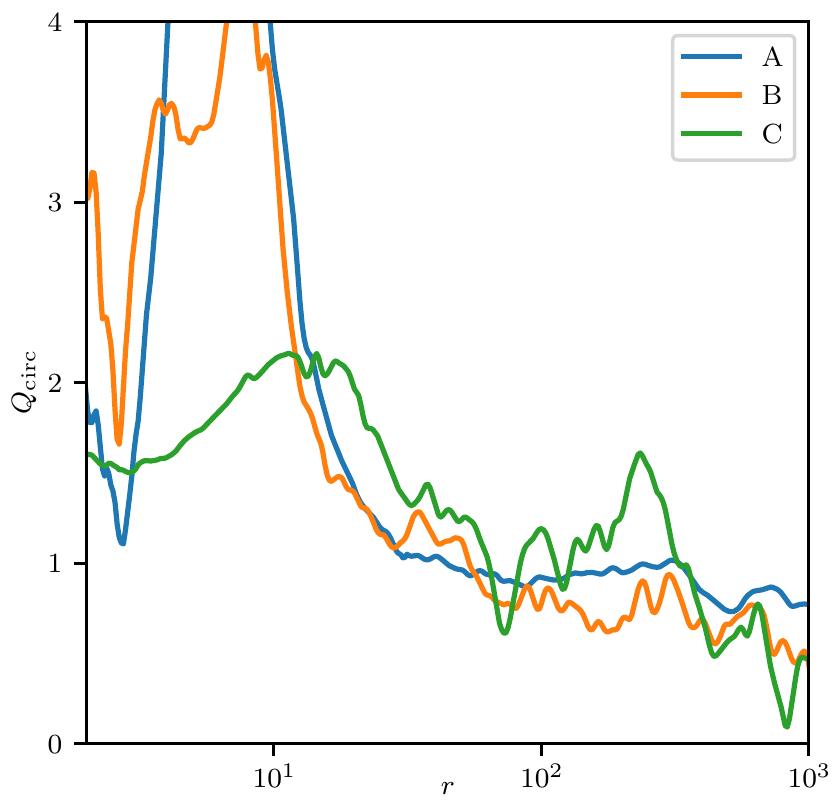}
  \caption{Meridional circulation criterion \citep{Philippov2017} calculated in the midplane for the fifth time chunk ($1.556\times10^5 \leq t < 2.2\times10^5$) of each simulation. Positive values predict a circulation pattern with outflow in the midplane. The values are positive everywhere, but they are roughly the same for all three simulations, whereas the simulations have different behaviors. \label{fig:circulation}}
\end{figure}

Classical convection is not the only way to achieve large-scale motions in a disk. A sufficiently small vertical stratification in entropy can lead to meridional circulation in which the material in the midplane of the disk moves outward, with inward flow through the coronal layers. Following the constant $\mdot$ case presented in \citet{Philippov2017} we define the circulation parameter
\begin{equation}
  \qcirc \equiv -\frac{\Gamma}{\delta_\rho + \delta_T} \paren[\bigg]{3 + \frac{3\Gamma + 2}{2\Gamma} \delta_T + \frac{1 - \Gamma}{\Gamma} \delta_\rho} - R^3 \paren[\bigg]{\frac{\pgas}{\rho}}^2 \frac{\pp^2s}{\pz^2},
\end{equation}
where $\delta_q$ denotes $\pp\log q/\pp\log R$. If this is positive we expect outflow in the midplane. $\qcirc$ is plotted for the same time chunks in Figure~\ref{fig:circulation}. In all three cases the values are positive throughout the disk. However, the values are similar for all three simulations; thus if this were the cause of the outflows in simulations~A and~C, we would expect it to lead to the same velocity structure in simulation~B, which is not seen.

\section{Comparison with Smaller Tori}
\label{sec:small}

Here we briefly mention two additional setups that are run for comparison. These two variants on the same small initial torus show how slight changes in the initial magnetic field can lead to different velocity structures in steady state.

In both cases we start with a \citet{Fishbone1976} torus around a nonspinning black hole with inner edge $r = 9$, pressure maximum $r = 17$, peak density $\rho = 1$, and adiabatic index $\Gamma = 4/3$. The grid extends inside the horizon and out to $r = 50$. At root level it contains $24^3$ cells, and we add three successive levels of mesh refinement, obtaining an effective resolution of $192^3$ within $37.5^\circ$ of the midplane.

The single-loop variant has a magnetic field initialized via
\begin{equation}
  A_\phi \propto \max(\rho - 0.2, 0),
\end{equation}
while the multiple loops variant employs \eqref{eq:a_phi} with $\pgasmin = 10^{-8}$, $\rmin = 10$, $\rmax = 35$, $\thetamin = 7\pi/18$, $\thetamax = 11\pi/18$, $N_r = 3$, and $N_\theta = 2$. In both cases the normalization is such that the density-weighted average of $\beta^{-1}$ is $10^{-2}$, and we perturb the initial state via \eqref{eq:perturb} with $\Delta_0 = 0.02$ and $k_R, k_z = 2\pi/5$.

The single-loop variant is comparable to standard tori widely used in the literature, for example when testing and demonstrating a new GRMHD code \citep{Gammie2003,Anton2006,White2016,Porth2017} or comparing such codes \citep{Porth2019}, though around a nonspinning black hole here. It is similar to a scaled-down version of simulation~B. The other small torus is more akin to a scaled-down version of simulation~A. The only difference between the two is the magnetic field configuration, which is chosen to remain SANE in both cases.

\begin{figure}
  \centering
  \includegraphics{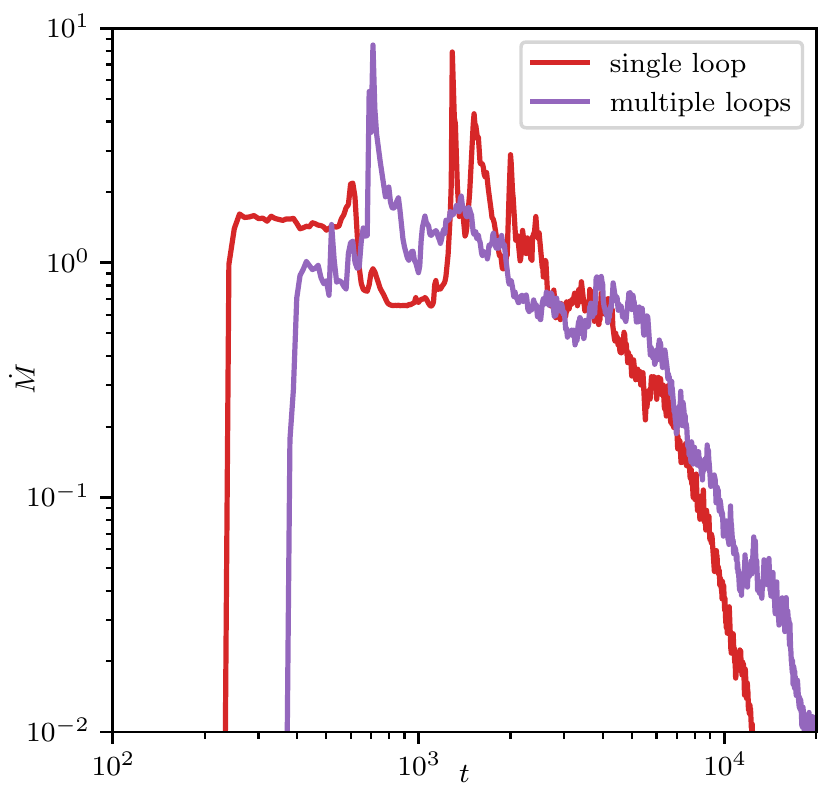}
  \caption{Horizon accretion rates as functions of time for the two small tori. The case with a single loop is a nonspinning analogue of the standard torus used to test and compare GRMHD codes as in \citet{Porth2019}. \label{fig:mdot_small}}
\end{figure}

Due to their small size, we can run these tori through steady state and even through depletion (by $t = 15{,}000$ for the single-loop case, slightly later with multiple loops) with relatively little computational expense. Figure~\ref{fig:mdot_small} shows the run of $\mdot$ with time in both cases.

\begin{figure}
  \centering
  \includegraphics{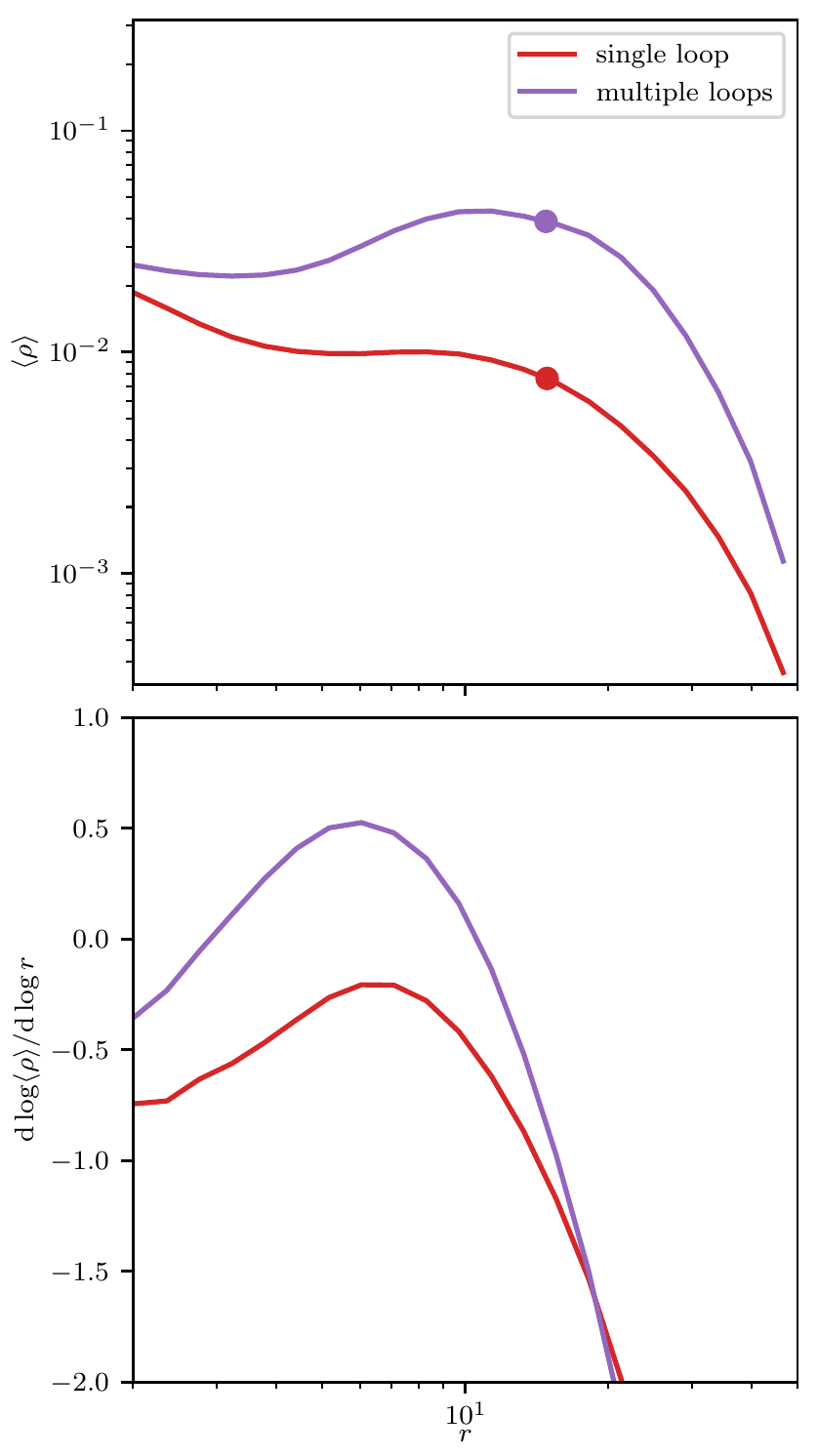}
  \caption{Radial density profiles and their power-law slopes for the small tori, averaged in time over $8000 \leq t \leq 10{,}000$. The points are at the same viscous radii as in Figure~\ref{fig:mdot_small}. The single-loop and multiple-loops variants are not too dissimilar from simulations~B and~A (their closest analogues) insofar as their slopes in the inner regions are shallower than the value of $-1$ seen in simulation~C. \label{fig:rho_r_small}}
\end{figure}

\begin{figure}
  \centering
  \includegraphics{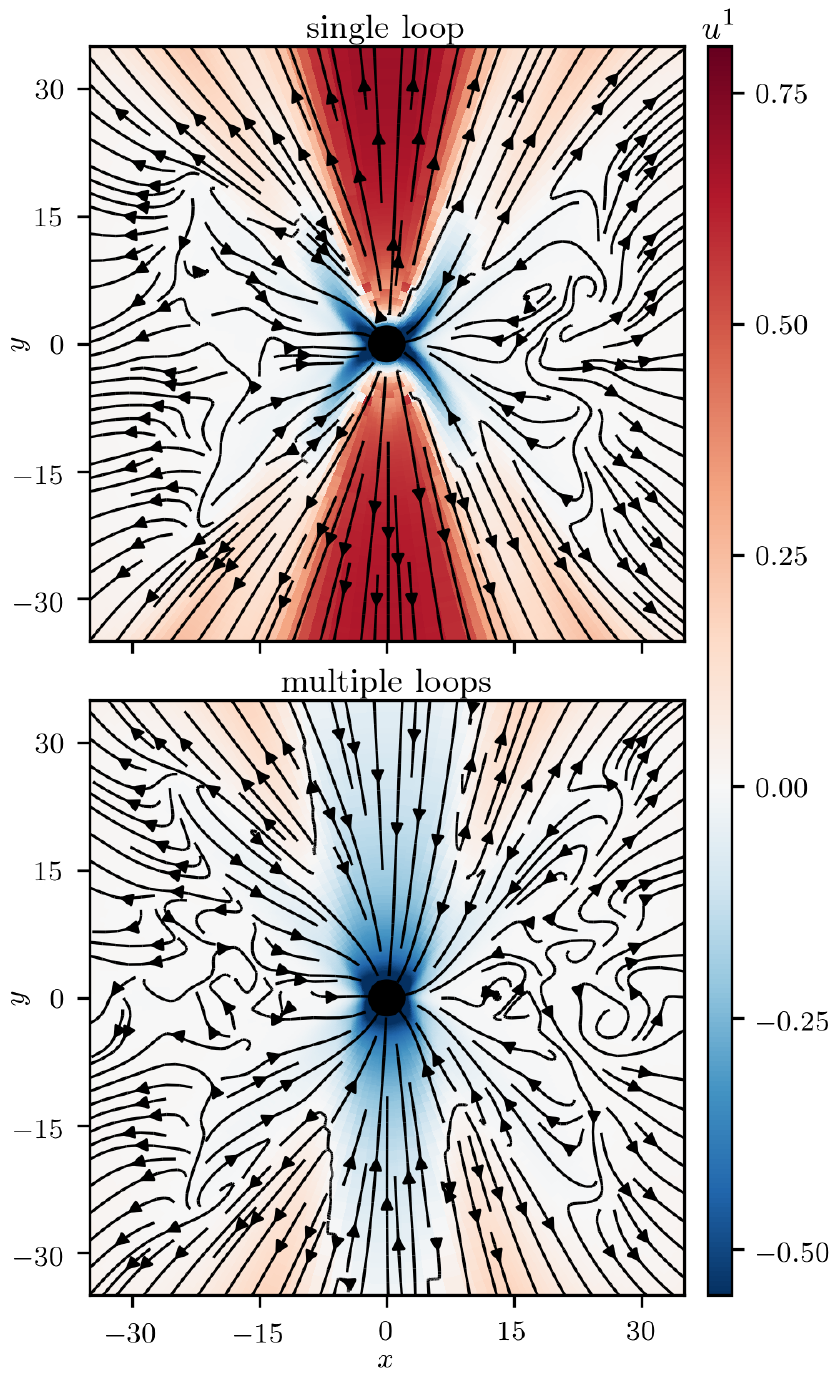}
  \caption{Average steady-state ($8000 \leq t \leq 10{,}000$) poloidal velocity fields for the small tori. As in Figure~\ref{fig:velocity}, background color highlights radial velocity. The single-loop and multiple-loops variants resemble simulations~B and~A, respectively, in the polar region. \label{fig:velocity_small}}
\end{figure}

Even with these widely used, standard, small tori, the final density and velocity structures depend on details of the initial field configuration. Density profiles, analogous to Figure~\ref{fig:rho_r}, are shown in Figure~\ref{fig:rho_r_small}, averaged in time over $8000 \leq t \leq 10{,}000$. Figure~\ref{fig:velocity_small} shows the corresponding velocity structure, analogous to Figure~\ref{fig:velocity}. With a single loop, a relatively strong polar outflow is driven, even though there is no black hole spin, nor is a MAD state obtained. This reflects the behavior of simulation~B, where some net vertical flux proves to be sufficient for launching polar outflows. On the other hand, the multiple-loops variant follows simulation~A in having polar inflow at all radii.

The midplane in the multiple-loops small torus does not experience the same bulk outflow as simulation~A. However, this can be attributed to larger \citeauthor{Fishbone1976} tori being less bound. The polar inflows in simulation~A could unbind most of the accretion disk, while the inner parts of the multiple-loops disk can still flow inward despite the liberation of gravitational potential energy nearby.

\bibliographystyle{aasjournal}
\bibliography{references}

\begin{thebibliography}{}
\expandafter\ifx\csname natexlab\endcsname\relax\def\natexlab#1{#1}\fi
\providecommand{\url}[1]{\href{#1}{#1}}
\providecommand{\dodoi}[1]{doi:~\href{http://doi.org/#1}{\nolinkurl{#1}}}
\providecommand{\doeprint}[1]{\href{http://ascl.net/#1}{\nolinkurl{http://ascl.net/#1}}}
\providecommand{\doarXiv}[1]{\href{https://arxiv.org/abs/#1}{\nolinkurl{https://arxiv.org/abs/#1}}}

\bibitem[{Ant{\'o}n {et~al.}(2006)Ant{\'o}n, Zanotti, Miralles, Mart{\'i},
  Ib{\'a}{\~n}ez, Font, \& Pons}]{Anton2006}
Ant{\'o}n, L., Zanotti, O., Miralles, J.~A., {et~al.} 2006, The Astrophysical
  Journal, 637, 296, \dodoi{10.1086/498238}

\bibitem[{Beckwith {et~al.}(2008)Beckwith, Hawley, \& Krolik}]{Beckwith2008}
Beckwith, K., Hawley, J.~F., \& Krolik, J.~H. 2008, The Astrophysical Journal,
  678, 1180, \dodoi{10.1086/533492}

\bibitem[{Begelman \& Volonteri(2017)}]{Begelman2017}
Begelman, M.~C., \& Volonteri, M. 2017, Monthly Notices of the Royal
  Astronomical Society, 464, 1102, \dodoi{10.1093/mnras/stw2446}

\bibitem[{Blandford \& Payne(1982)}]{Blandford1982}
Blandford, R., \& Payne, D. 1982, Monthly Notices of the Royal Astronomical
  Society, 199, 883, \dodoi{10.1093/mnras/199.4.883}

\bibitem[{Blandford \& Znajek(1977)}]{Blandford1977}
Blandford, R., \& Znajek, R. 1977, Monthly Notices of the Royal Astronomical
  Society, 179, 433, \dodoi{10.1093/mnras/179.3.433}

\bibitem[{Blandford \& Begelman(1999)}]{Blandford1999}
Blandford, R.~D., \& Begelman, M.~C. 1999, Monthly Notices of the Royal
  Astronomical Society, 303, L1, \dodoi{10.1046/j.1365-8711.1999.02358.x}

\bibitem[{Bondi(1952)}]{Bondi1952}
Bondi, H. 1952, Monthly Notices of the Royal Astronomical Society, 112, 195,
  \dodoi{10.1093/mnras/112.2.195}

\bibitem[{Chael {et~al.}(2019)Chael, Narayan, \& Johnson}]{Chael2019}
Chael, A., Narayan, R., \& Johnson, M.~D. 2019, Monthly Notices of the Royal
  Astronomical Society, 486, 2873, \dodoi{10.1093/mnras/stz988}

\bibitem[{Chael {et~al.}(2018)Chael, Rowan, Narayan, Johnson, \&
  Sironi}]{Chael2018}
Chael, A., Rowan, M., Narayan, R., Johnson, M., \& Sironi, L. 2018, Monthly
  Notices of the Royal Astronomical Society, 478, 5209,
  \dodoi{10.1093/mnras/sty1261}

\bibitem[{Chakrabarti(1985)}]{Chakrabarti1985}
Chakrabarti, S.~K. 1985, The Astrophysical Journal, 288, 1,
  \dodoi{10.1086/162755}

\bibitem[{Dexter {et~al.}(2014)Dexter, Kelly, Bower, Marrone, Stone, \&
  Plambeck}]{Dexter2014}
Dexter, J., Kelly, B., Bower, G.~C., {et~al.} 2014, Monthly Notices of the
  Royal Astronomical Society, 442, 2797, \dodoi{10.1093/mnras/stu1039}

\bibitem[{Dexter {et~al.}(2012)Dexter, McKinney, \& Agol}]{Dexter2012}
Dexter, J., McKinney, J.~C., \& Agol, E. 2012, Monthly Notices of the Royal
  Astronomical Society, 421, 1517, \dodoi{10.1111/j.1365-2966.2012.20409.x}

\bibitem[{Doeleman {et~al.}(2008)Doeleman, Weintroub, Rogers, Plambeck, Freund,
  Tilanus, Friberg, Ziurys, Moran, Corey, Young, Smythe, Titus, Marrone,
  Cappallo, Bock, Bower, Chamberlin, Davis, Krichbaum, Lamb, Maness, Niell,
  Roy, Strittmatter, Werthimer, Whitney, \& Woody}]{Doeleman2008}
Doeleman, S.~S., Weintroub, J., Rogers, A.~E., {et~al.} 2008, Nature, 455, 78,
  \dodoi{10.1038/nature07245}

\bibitem[{Fishbone \& Moncrief(1976)}]{Fishbone1976}
Fishbone, L.~G., \& Moncrief, V. 1976, The Astrophysical Journal, 207, 962,
  \dodoi{10.1086/154565}

\bibitem[{Gammie {et~al.}(2003)Gammie, McKinney, \& T{\'o}th}]{Gammie2003}
Gammie, C.~F., McKinney, J.~C., \& T{\'o}th, G. 2003, The Astrophysical
  Journal, 589, 444, \dodoi{10.1086/374594}

\bibitem[{Hawley \& Balbus(2002)}]{Hawley2002}
Hawley, J.~F., \& Balbus, S.~A. 2002, The Astrophysical Journal, 573, 738,
  \dodoi{10.1086/340765}

\bibitem[{Ichimaru(1977)}]{Ichimaru1977}
Ichimaru, S. 1977, The Astrophysical Journal, 214, 840, \dodoi{10.1086/155314}

\bibitem[{Igumenshchev \& Abramowicz(2000)}]{Igumenshchev2000}
Igumenshchev, I.~V., \& Abramowicz, M.~A. 2000, The Astrophysical Journal
  Supplement Series, 130, 463, \dodoi{10.1086/317354}

\bibitem[{Koz{\l}owski {et~al.}(1978)Koz{\l}owski, Jaroszy{\'n}ski, \&
  Abramowicz}]{Kozlowski1978}
Koz{\l}owski, M., Jaroszy{\'n}ski, M., \& Abramowicz, M. 1978, Astronomy \&
  Astrophysics, 63, 209

\bibitem[{Mignone(2014)}]{Mignone2014}
Mignone, A. 2014, Journal of Computational Physics, 270, 784,
  \dodoi{10.1016/j.jcp.2014.04.001}

\bibitem[{Mignone {et~al.}(2009)Mignone, Ugliano, \& Bodo}]{Mignone2009}
Mignone, A., Ugliano, M., \& Bodo, G. 2009, Monthly Notices of the Royal
  Astronomical Society, 393, 1141, \dodoi{10.1111/j.1365-2966.2008.14221.x}

\bibitem[{Mo{\'s}cibrodzka {et~al.}(2017)Mo{\'s}cibrodzka, Dexter, Davelaar, \&
  Falcke}]{Moscibrodzka2017}
Mo{\'s}cibrodzka, M., Dexter, J., Davelaar, J., \& Falcke, H. 2017, Monthly
  Notices of the Royal Astronomical Society, 468, 2214,
  \dodoi{10.1093/mnras/stx587}

\bibitem[{Mo{\'s}cibrodzka {et~al.}(2016)Mo{\'s}cibrodzka, Falcke, \&
  Shiokawa}]{Moscibrodzka2016}
Mo{\'s}cibrodzka, M., Falcke, H., \& Shiokawa, H. 2016, Astronomy \&
  Astrophysics, 586, A38, \dodoi{10.1051/0004-6361/201526630}

\bibitem[{Mo{\'s}cibrodzka {et~al.}(2014)Mo{\'s}cibrodzka, Falcke, Shiokawa, \&
  Gammie}]{Moscibrodzka2014}
Mo{\'s}cibrodzka, M., Falcke, H., Shiokawa, H., \& Gammie, C.~F. 2014,
  Astronomy \& Astrophysics, 570, A7, \dodoi{10.1051/0004-6361/201424358}

\bibitem[{Narayan {et~al.}(2000)Narayan, Igumenshchev, \&
  Abramowicz}]{Narayan2000}
Narayan, R., Igumenshchev, I.~V., \& Abramowicz, M.~A. 2000, The Astrophysical
  Journal, 539, 798, \dodoi{10.1086/309268}

\bibitem[{Narayan {et~al.}(2003)Narayan, Igumenshchev, \&
  Abramowicz}]{Narayan2003}
---. 2003, Publications of the Astronomical Society of Japan, 55, L69,
  \dodoi{10.1093/pasj/55.6.L69}

\bibitem[{Narayan {et~al.}(1998)Narayan, Mahadevan, Grindlay, Popham, \&
  Gammie}]{Narayan1998}
Narayan, R., Mahadevan, R., Grindlay, J.~E., Popham, R.~G., \& Gammie, C. 1998,
  The Astrophysical Journal, 492, 554, \dodoi{10.1086/305070}

\bibitem[{Narayan {et~al.}(2012)Narayan, S{\k{a}}dowski, Penna, \&
  Kulkarni}]{Narayan2012}
Narayan, R., S{\k{a}}dowski, A., Penna, R.~F., \& Kulkarni, A.~K. 2012, Monthly
  Notices of the Royal Astronomical Society, 426, 3241,
  \dodoi{10.1111/j.1365-2966.2012.22002.x}

\bibitem[{Narayan \& Yi(1994)}]{Narayan1994}
Narayan, R., \& Yi, I. 1994, The Astrophysical Journal, 428, L13,
  \dodoi{10.1086/187381}

\bibitem[{Narayan \& Yi(1995)}]{Narayan1995}
---. 1995, The Astrophysical Journal, 452, 710, \dodoi{10.1086/176343}

\bibitem[{Noble {et~al.}(2007)Noble, Leung, Gammie, \& Book}]{Noble2007}
Noble, S.~C., Leung, P.~K., Gammie, C.~F., \& Book, L.~G. 2007, Classical and
  Quantum Gravity, 24, S259, \dodoi{10.1088/0264-9381/24/12/S17}

\bibitem[{Penna {et~al.}(2013)Penna, Kulkarni, \& Narayan}]{Penna2013}
Penna, R.~F., Kulkarni, A., \& Narayan, R. 2013, Astronomy \& Astrophysics,
  559, A116, \dodoi{10.1051/0004-6361/201219666}

\bibitem[{Philippov \& Rafikov(2017)}]{Philippov2017}
Philippov, A.~A., \& Rafikov, R.~R. 2017, The Astrophysical Journal, 837, 101,
  \dodoi{10.3847/1538-4357/aa60ca}

\bibitem[{Porth {et~al.}(2017)Porth, Olivares, Mizuno, Younsi, Rezzolla,
  Moscibrodzka, Falcke, \& Kramer}]{Porth2017}
Porth, O., Olivares, H., Mizuno, Y., {et~al.} 2017, Computational Astrophysics
  and Cosmology, 4, 1, \dodoi{10.1186/s40668-017-0020-2}

\bibitem[{Porth {et~al.}(2019)Porth, Chatterjee, Narayan, Gammie, Mizuno,
  Anninos, Baker, Bugli, kwan Chan, Davelaar, {Del~Zanna}, Etienne, Fragile,
  Kelly, Liska, Markoff, McKinney, Mishra, Noble, Olivares, Prather, Rezzolla,
  Ryan, Stone, Tomei, White, Younsi, \& {The Event Horizon Telescope
  Collaboration}}]{Porth2019}
Porth, O., Chatterjee, K., Narayan, R., {et~al.} 2019, The Astrophysical
  Journal Supplement Series, 243, 26, \dodoi{10.3847/1538-4365/ab29fd}

\bibitem[{Quataert \& Gruzinov(2000)}]{Quataert2000}
Quataert, E., \& Gruzinov, A. 2000, The Astrophysical Journal, 539, 809,
  \dodoi{10.1086/309267}

\bibitem[{Rees {et~al.}(1982)Rees, Begelman, Blandford, \& Phinney}]{Rees1982}
Rees, M., Begelman, M., Blandford, R., \& Phinney, E. 1982, Nature, 295, 17,
  \dodoi{10.1038/295017a0}

\bibitem[{Ressler {et~al.}(2018)Ressler, Quataert, \& Stone}]{Ressler2018}
Ressler, S., Quataert, E., \& Stone, J. 2018, Monthly Notices of the Royal
  Astronomical Society, 478, 3544, \dodoi{10.1093/mnras/sty1146}

\bibitem[{Ressler {et~al.}(2017)Ressler, Tchekhovskoy, Quataert, \&
  Gammie}]{Ressler2017}
Ressler, S., Tchekhovskoy, A., Quataert, E., \& Gammie, C. 2017, Monthly
  Notices of the Royal Astronomical Society, 467, 3604,
  \dodoi{10.1093/mnras/stx364}

\bibitem[{Ryan {et~al.}(2018)Ryan, Ressler, Dolence, Gammie, \&
  Quataert}]{Ryan2018}
Ryan, B.~R., Ressler, S.~M., Dolence, J.~C., Gammie, C., \& Quataert, E. 2018,
  The Astrophysical Journal, 864, 126, \dodoi{10.3847/1538-4357/aad73a}

\bibitem[{Shakura \& Sunyaev(1973)}]{Shakura1973}
Shakura, N., \& Sunyaev, R. 1973, Astronomy \& Astrophysics, 24, 337

\bibitem[{Stone {et~al.}(1999)Stone, Pringle, \& Begelman}]{Stone1999}
Stone, J.~M., Pringle, J.~E., \& Begelman, M.~C. 1999, Monthly Notices of the
  Royal Astronomical Society, 310, 1002,
  \dodoi{10.1046/j.1365-8711.1999.03024.x}

\bibitem[{Tchekhovskoy {et~al.}(2011)Tchekhovskoy, Narayan, \&
  McKinney}]{Tchekhovskoy2011}
Tchekhovskoy, A., Narayan, R., \& McKinney, J.~C. 2011, Monthly Notices of the
  Royal Astronomical Society, 418, L79,
  \dodoi{10.1111/j.1745-3933.2011.01147.x}

\bibitem[{{The Event Horizon Telescope
  Collaboration}(2019{\natexlab{a}})}]{EHT2019a}
{The Event Horizon Telescope Collaboration}. 2019{\natexlab{a}}, The
  Astrophysical Journal Letters, 875, L1, \dodoi{10.3847/2041-8213/ab0ec7}

\bibitem[{{The Event Horizon Telescope
  Collaboration}(2019{\natexlab{b}})}]{EHT2019b}
---. 2019{\natexlab{b}}, The Astrophysical Journal Letters, 875, L5,
  \dodoi{10.3847/2041-8213/ab0f43}

\bibitem[{{The Event Horizon Telescope
  Collaboration}(2019{\natexlab{c}})}]{EHT2019c}
---. 2019{\natexlab{c}}, The Astrophysical Journal Letters, 875, L6,
  \dodoi{10.3847/2041-8213/ab1141}

\bibitem[{White {et~al.}(2016)White, Stone, \& Gammie}]{White2016}
White, C.~J., Stone, J.~M., \& Gammie, C.~F. 2016, The Astrophysical Journal
  Supplement Series, 225, 22, \dodoi{10.3847/0067-0049/225/2/22}

\bibitem[{Yuan {et~al.}(2012)Yuan, Wu, \& Bu}]{Yuan2012}
Yuan, F., Wu, M., \& Bu, D. 2012, The Astrophysical Journal, 761, 129,
  \dodoi{10.1088/0004-637X/761/2/129}

\end{thebibliography}

\end{document}